\documentclass[fleqn,usenatbib,useAMS]{mnras}

\usepackage{newtxtext,newtxmath}

\usepackage[T1]{fontenc}
\usepackage{ae,aecompl}

\usepackage{graphicx}	
\usepackage{amsmath}	
\usepackage{amssymb}	
\usepackage[flushleft]{threeparttable}
\usepackage{deluxetable}
\usepackage{booktabs}
\usepackage{soul}

\newcommand{\mstar}{\rm{M}$_{\ast}$}
\newcommand{\msun}{\rm{M}$_{\odot}$}

\makeatletter
\def\@to{to}
\makeatother

\title[Modelling chemical abundance distributions for dwarf galaxies]{Modelling chemical abundance distributions for dwarf galaxies in the Local Group: the impact of turbulent metal diffusion}

\author[I. Escala et al.]{
Ivanna Escala,$^{1}$\thanks{E-mail: ie@astro.caltech.edu}
Andrew Wetzel,$^{2,3,4}$
Evan N. Kirby,$^{1}$
Philip F. Hopkins,$^{2}$
\newauthor{
Xiangcheng Ma,$^{2}$
Coral Wheeler,$^{2}$
Du\v{s}an Kere\v{s},$^{5}$
Claude-Andr\'{e} Faucher-Gigu\`{e}re,$^{6}$
}
\newauthor{
Eliot Quataert$^{7}$
}
\\
$^{1}$Department of Astronomy, California Institute of Technology, Pasadena, CA, USA\\
$^{2}$TAPIR, California Institute of Technology, Pasadena, CA USA\\
$^{3}$The Observatories of the Carnegie Institution for Science, Pasadena, CA, USA\\
$^{4}$Department of Physics, University of California, Davis, CA\\
$^{5}$Department of Physics, Center for Astrophysics and Space Science, University of California, San Diego, La Jolla, CA USA\\
$^{6}$Department of Physics and Astronomy and CIERA, Northwestern University, Evanston, IL, USA\\
$^{7}$Department of Astronomy and Theoretical Astrophysics Center, University of California, Berkeley, CA, USA}

\date{Accepted 2017 November 1. Received 2017 October 31; in original form 2017 October 17}

\pubyear{2017}

\begin{document}
\label{firstpage}
\pagerange{\pageref{firstpage}--\pageref{lastpage}}
\maketitle

\begin{abstract}
We investigate stellar metallicity distribution functions (MDFs), including Fe and $\alpha$-element abundances, in dwarf galaxies from the Feedback in Realistic Environments (FIRE) project. We examine both isolated dwarf galaxies and those that are satellites of a Milky Way-mass galaxy. In particular, we study the effects of including a sub-grid turbulent model for the diffusion of metals in gas. Simulations that include diffusion have narrower MDFs and abundance ratio distributions, because diffusion drives individual gas and star particles toward the average metallicity. This effect provides significantly better agreement with observed abundance distributions in dwarf galaxies in the Local Group, including small intrinsic scatter in [$\alpha$/Fe] vs. [Fe/H] of $\lesssim$ 0.1 dex. This small intrinsic scatter arises in our simulations because the interstellar medium in dwarf galaxies is well-mixed at nearly all cosmic times, such that stars that form at a given time have similar abundances to $\lesssim$ 0.1 dex. Thus, most of the scatter in abundances at $z$ = 0 arises from redshift evolution and not from instantaneous scatter in the ISM. We find similar MDF widths and intrinsic scatter for satellite and isolated dwarf galaxies, which suggests that environmental effects  play a minor role compared with internal chemical evolution in our simulations. Overall, with the inclusion of metal diffusion, our simulations reproduce abundance distribution widths of observed low-mass galaxies, enabling detailed studies of chemical evolution in galaxy formation.
\end{abstract}

\begin{keywords}
methods: numerical -- diffusion -- galaxies: abundances -- galaxies: dwarf 
\end{keywords}



\section{Introduction}

Dwarf galaxies, which probe the low-mass (\mstar \ $\lesssim$ 10$^{9}$ \msun) end of the galaxy mass spectrum, serve as an environment to test models of galaxy formation. Dwarf galaxies are extremely sensitive to feedback effects from supernovae (SNe) and stellar winds, owing to their shallow gravitational potential wells.  This results in significant mass loss \citep[e.g.,][]{Dekel1986TheFormation}, and thus metal loss, which is relevant for studies of galactic chemical evolution. 

Owing to their small sizes, low masses, and relatively inefficient star formation, dwarf galaxies can be challenging to simulate accurately. However, given that they form out of a small volume, dwarf galaxies are ideal for targeted zoom-in simulations. Hydrodynamical simulations of galaxy evolution have achieved increasingly high baryonic particle mass resolution, such that only a few SNe occur per star particle, where each star particle represents a single stellar population. The limited mass-sampling of chemical enrichment histories thus becomes important for simulating dwarf galaxies in the given stellar mass range. Consequently, the predicted chemical evolution of dwarf galaxies will be impacted by the specific feedback implementation used in such simulations. For example, the predicted abundances are subject to stochastic sampling of nucleosynthetic events (e.g., the ``enrichment sampling problem'' of \citealt{Wiersma2009}). 

A majority of the studies investigating the detailed properties of chemical evolution have been based on one-zone numerical models of galactic chemical evolution with instantaneous mixing \citep{Lanfranchi2003, Lanfranchi2007,Lanfranchi2010, Lanfranchi2006,Kirby2011a}.
Only recently have hydrodynamical simulations of cosmological isolated dwarf galaxies attempted to go beyond accurately reproducing the stellar mass-metallicity relation \citep{Ma2016} to include more detailed chemical evolution properties \citep{Marcolini2008,Revaz2009,Sawala2010,RevazJablonka2012}. For example, \citet{Sawala2010} simulated the metallicity distribution functions (MDFs) and $\alpha$-element abundance ratios for isolated dwarf galaxies, yet the MDF is broad compared to observations \citep{Kirby2011a} and contains a pronounced, unobserved low-metallicity tail, whereas the scatter in the $\alpha$-element abundances is large, particularly at low-metallicity \citep{Tolstoy2009, Kirby2011b,FrebelNorris2015}. However, both MDFs and abundance ratios can serve as tests of the metal injection scheme, energy and momentum injection from stellar feedback, yields, and microphysics in the interstellar medium (ISM). 

Until recently (e.g., \citealt{Shen2010,Shen2013,Pilkington2012,Brook2014}), sub-grid turbulent diffusion has been neglected in many astrophysical simulations based on Lagrangian methods, such as smoothed particle hydrodynamics (SPH) or mesh-free methods. This is despite the fact that supersonic, compressible flows result in a turbulent cascade that transports momentum to small scales, where viscous forces begin to dominate \citep{Wadsley2008}. Furthermore, ``standard'' SPH methods are known to suppress mixing on small scales, in contrast to Eulerian codes \citep{Agertz2007} and other finite-volume methods \citep{Hopkins2015}, which include intrinsic numerical diffusion (e.g., \citealt{Recchi2001}). In contrast to most prior galaxy evolution simulations, the Feedback in Realistic Environments (FIRE)\footnote{The FIRE project website is \url{http://fire.northwestern.edu}.} project \citep{Hopkins2014,Hopkins2017arXiv} has recently implemented \citep{Su2017} a model for turbulent metal diffusion (TMD) due to unresolved, small-scale eddies.

\citet{Su2017} considered the impact of sub-grid metal diffusion in the context of a realistic, multi-phase ISM influenced by strong stellar feedback processes. They found that sub-grid metal diffusion does not significantly impact cooling physics, having no systematic effect on galactic star formation rates. This is in contrast to the findings of \citet{Shen2010} and \citet{Pilkington2012} on the effects of including fluid microphysics. For example, \citet{Shen2010} concluded that simulations without sub-grid metal diffusion produce slightly fewer stars, since fewer gas particles experience gas cooling and subsequently turn into stars. Comparatively low spatial resolution, such that the turbulent driving scales are not resolved, could potentially explain the discrepancy \citep{Su2017}. 
Although sub-grid metal diffusion does not significantly impact cooling rates or star formation rates in the FIRE simulations, sub-grid metal diffusion is expected to strongly impact chemical evolution.

Motivated by previous studies (e.g., \citealt{Aguirre2005}) that have shown that metals are too inhomogeneous in simulations, the introduction of a diffusive term on sub-grid scales has recently been explored as a promising solution to the problem of reproducing realistic MDFs and abundances. \citet{Shen2010} implemented a turbulence-induced mixing model in SPH simulations based on velocity shear, as opposed to the velocity-dispersion based model of \citet{Greif2009}. Using the latter model, \citet{Jeon2017arXiv} incorporated metal diffusion into a fully cosmological study of the chemical abundances of ultra-faint dwarf galaxies. \citet{Williamson2016} investigated sub-grid metal mixing in non-cosmological, isolated dwarf galaxies, and found that the metallicity of stars is not strongly dependent on how the diffusivity is calculated from the the velocity distribution. In addition, they observed a reduction of scatter in stellar abundances and the suppression of low-metallicity star formation. \citet{Pilkington2012} found that sub-grid metal diffusion reduced the overproduction of extremely metal poor stars, except for M33-like spiral galaxies, as opposed to dwarf galaxies. 

\citet{HiraiSaitoh2017} explored the efficiency of sub-grid metal mixing in non-cosmological isolated dwarf galaxy simulations, focusing on reproducing the scatter in barium inferred from extremely metal-poor stars \citep{Suda2008}. They concluded that the timescale for metal mixing necessary to reproduce observations of barium is $\lesssim$ 40 Myr, which is shorter than the typical dynamical timescale of dwarf galaxies ($\sim$ 100 Myr). \citet{Kawata2014} investigated the impact of strong stellar feedback in a simulation of a WLM-like, non-cosmological dwarf disc galaxy, finding that including sub-grid diffusion maintained low-metallicity in star-forming regions owing to efficient metal mixing in the ISM. Comparing their results on the dispersion in the abundances of newly formed stars to observations, they concluded that their sub-grid diffusion was likely too strong. \citet{Revaz2016} showed that metal diffusion can reproduce the low scatter in $\alpha$-elements at low metallicity in particle-based simulations. However, they concluded that a ``smoothed metallicity scheme'' \citep{Wiersma2009}, in which metallicity is treated as a smoothly varying function and involves no explicit redistribution of metals, is preferable over sub-grid metal diffusion to reproduce the observed dispersion in abundances of dwarf galaxies. 

In this work, we use the high-resolution, cosmological zoom-in simulations of the FIRE project to analyse the impact of turbulent metal diffusion on observables related to chemical evolution for simulated dwarf galaxies in the mass range \mstar($z$ = 0) $\sim$ 7 $\times$ 10$^{5}$ - 2 $\times$ 10$^{8}$ \msun. First, we study a small sample of cosmological field dwarf galaxies simulated at very high resolution, then expand our analysis to include satellite and isolated dwarf galaxies of a Milky Way (MW) mass halo \citep{Wetzel2016}. We find that the inclusion of a physically-motivated, sub-grid turbulent diffusion model produces MDFs and abundance ratios consistent with observations of Local Group (LG) dwarf galaxies. We confirm the necessity of including sub-grid metal mixing in Lagrangian hydrodynamical codes, while taking into account a multi-phase ISM, explicit stellar and radiative feedback, the impact of cosmological accretion, and environmental effects. 

\vspace{-0.4cm}
\section{Simulations}

\begin{table}
\caption{FIRE Simulation Properties\label{tab:sims}}
\begin{threeparttable}
\begin{tabular*}{\columnwidth}{l @{\extracolsep{\fill}} ccc}
	\toprule
    Simulation\tnote{a} & $M_{\ast}$\tnote{b} (10$^{6}$ $M_{\odot}$) &
    $\langle$[Fe/H]$\rangle$\tnote{c} (dex) & $\sigma$\tnote{d} (dex) \\ \midrule
 	\textbf{m10q} & 1.7 & $-$2.16 & 0.55 \\
	\textbf{m10q.md} & 2.0 & $-$2.12 & 0.41 \\
	\textbf{m10v}\footnotemark & 1.1 & $-$1.82 & 0.52 \\
	\textbf{m10v.md} & 1.5 & $-$1.54 & 0.34 \\ \bottomrule
\end{tabular*}
\begin{tablenotes}
\item Note. \textemdash \ All quantities are determined at $z$ = 0. For all isolated dwarf galaxy simulations, the star particle spatial resolution $h$ is 1.4 pc and the mass resolution is 250 $M_{\odot}$. For more details on the methods used to simulate the cosmological, isolated dwarf galaxies, see \citet{Onorbe2014}, \citet{Hopkins2014}, and \citet{Hopkins2017arXiv}.
\item[a] The simulation naming convention reflects the halo mass, e.g., m10 $\implies$ $M_{\textrm{halo}}$ $\sim$ 10$^{10}$ $M_{\odot}$ at $z$ = 0. The designations ``q" and ``v" reflect initial conditions that distinguish between halos dominated by early- and late-time star formation respectively.  The addition  ``.md" indicates that the simulation was run with sub-grid turbulent metal diffusion  (\S~\ref{sec:tmd}).
\item[b] Defined as the mass within the radius that contains 90\% of the stellar mass, $r_{90}$.
\item[c] The mass-weighted average metallicity of star particles within $r_{90}$ (Eq~\ref{eq:metal_def_ivanna}) and,
\item[d] the associated standard deviation, or intrinsic spread in the metallicity distribution.
\end{tablenotes}
\end{threeparttable}
\end{table}
\footnotetext{The galaxy labelled \textbf{m10v} analysed in this paper (and the version of the run with sub-grid metal diffusion, \textbf{m10v.md}) is not the usual \textbf{m10v} included in previous FIRE papers. Instead, this \textbf{m10v} is a significantly contaminated galaxy in the same cosmological volume with a larger stellar mass (1.1 $\times$ $10^6$ \msun, as compared to $\sim$ 10$^{5}$ \msun) at the outskirts of the high-resolution region. We emphasize that the comparisons between our \textbf{m10v} and \textbf{m10v.md} are internally consistent, since they both suffer from low-resolution dark matter contamination, and that despite this, they still produce realistic dwarf galaxies. Additionally, the majority of our conclusions are based on the properties of the collective simulated dwarf galaxy sample, particularly \textbf{m10q} and the satellite and isolated dwarf galaxies around \textbf{m12i} (\S~\ref{sec:latte})}

We present a generalized summary of the relevant features of the FIRE simulations \citep{Hopkins2014,Hopkins2017arXiv}. The simulations analysed here (e.g., Table~\ref{tab:sims}) were run with the ``FIRE-2'' \citep{Hopkins2017arXiv} rerun of GIZMO\footnote{The public version of GIZMO is available at \url{http://www.tapir.caltech.edu/~phopkins/Site/GIZMO.html}} in its Meshless Finite Mass (MFM) mode \citep{Hopkins2015}. All feedback quantities are based on stellar evolution models that are identical between ``FIRE-1'' \citep{Hopkins2014} and ``FIRE-2'', such that galaxy-scale properties do not qualitatively change between versions of the code \citep{Hopkins2017arXiv}.

The MFM method combines advantages from both SPH and grid-based codes. FIRE cosmological simulations of dwarf galaxies have reproduced several key observations. These include the bursty star formation and outflows generated by low-mass galaxies at high redshift \citep{Muratov2015}, the stellar mass-halo mass relation at both $z$ = 0 \citep{Onorbe2015} and at high redshift \citep{Ma2017a}, the dark matter halo profile in dwarf galaxies \citep{Chan2015}, the size evolution and age gradients of dwarf galaxies \citep{ElBadry2016}, the stellar mass-metallicity relation \citep{Ma2016}, and the stellar kinematics of dwarf galaxies \citep{Wheeler2017}, all without calibration to match observations at $z$ = 0. 


\vspace{-0.3cm}
\subsection{Gas Cooling, Star Formation, and Feedback}
\label{sec:gas_cool_sf}

Gas follows a cooling curve from 10 -- 10$^{10}$ K, 
with cooling at low temperatures due to molecular transitions and metal-line fine structure transitions, and primordial and metal line cooling at higher temperatures ($\geq$ 10$^{4}$ K). A uniform, redshift-dependent photoionizing background \citep{FaucherGiguere2009} is taken into account at each timestep when determining the cooling rates. 

Star formation occurs only in dense, molecular, self-gravitating regions with $n_{\textrm{crit}} >$ 1000 cm$^{-3}$. When these conditions are met, stars form at 100\% efficiency per local free-fall time, although stellar feedback rapidly regulates the global star formation efficiency to a few percent on the scales of giant molecular clouds \citep{Orr2016WhatFeedback}. The newly formed star particle inherits its metallicity from its progenitor gas particle. 11 total chemical species (H, He, C, N, O, Ne, Mg, Si, S, Ca, Fe), including $\alpha$- and Fe peak elements, which are particularly relevant to constraining star formation history (SFH), are tracked  in addition to the total metallicity \citep{Wiersma2009}. Each star particle is treated as a single stellar population with a \citet{Kroupa2002} initial mass function (IMF), with known age, mass, and metallicity. Feedback quantities such as luminosity, Type II SN rates, and the rates of mass and metal loss are calculated based on stellar population models (STARBURST99; \citealt{Leitherer1999}). SN explosions occur discretely, as opposed to modelling their collective effects. Metal yields for Type Ia SNe are adopted from \citet{Iwamoto1999}, where the rates follow \citet{Mannucci2006}, including both prompt and delayed populations. Metal yields for Type II SNe \citep{Nomoto2006} and  stellar winds (AGB \& O-stars) are also included, as well as their contributions to ejecta energy, momentum, and mass. All feedback quantities are deposited directly into the ISM (gas particles) surrounding a given star particle, where mass, energy, and momentum are conserved. Radiative feedback from local photo-ionization, photo-electric heating, and radiation pressure, are also included.

\vspace{-0.4cm}
\subsection{Turbulent Metal Diffusion}
\label{sec:tmd}

\begin{figure}
\centering
\includegraphics[width=\columnwidth]{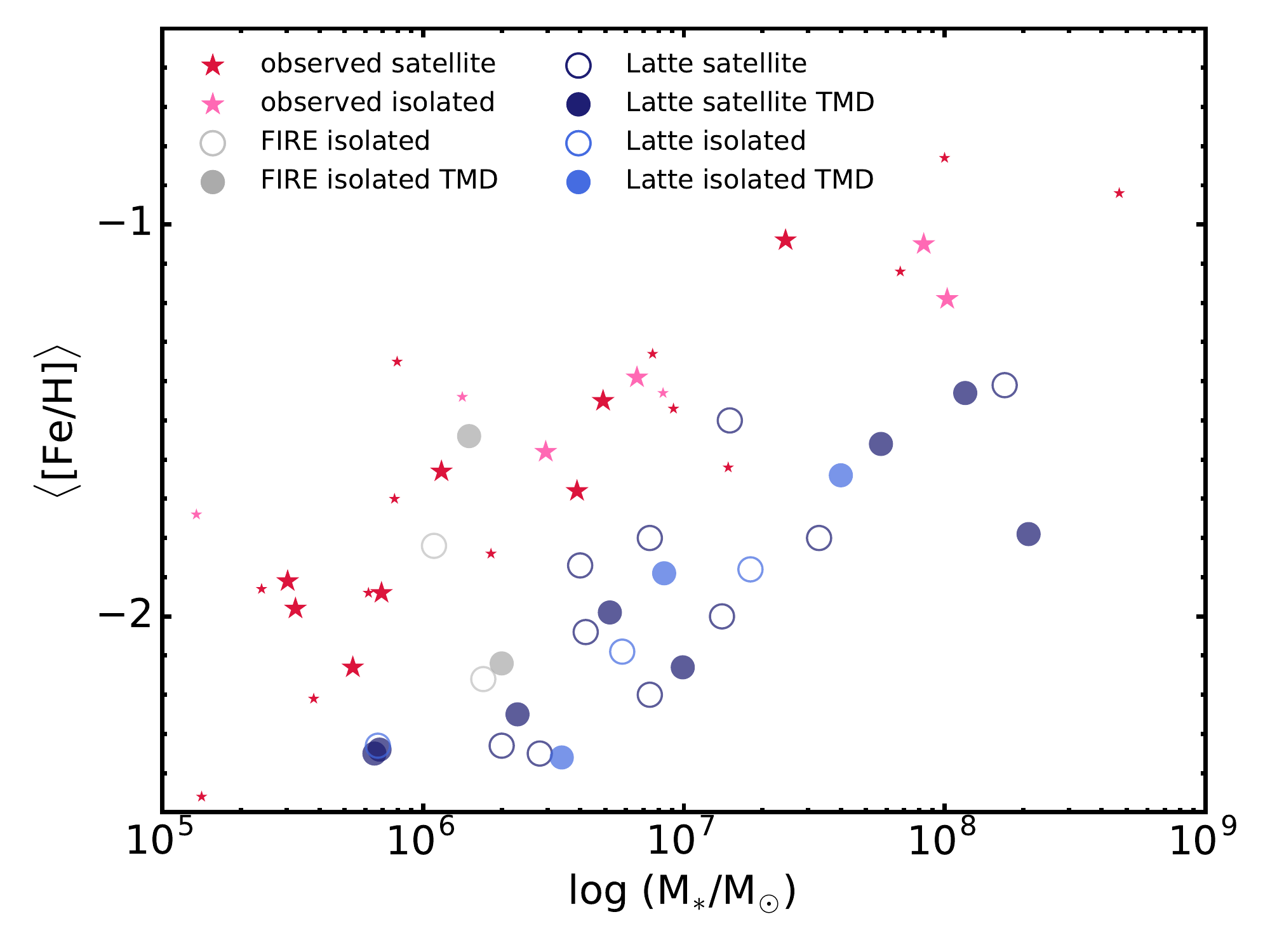}
\hspace{-0.5cm}
\includegraphics[width=\columnwidth]{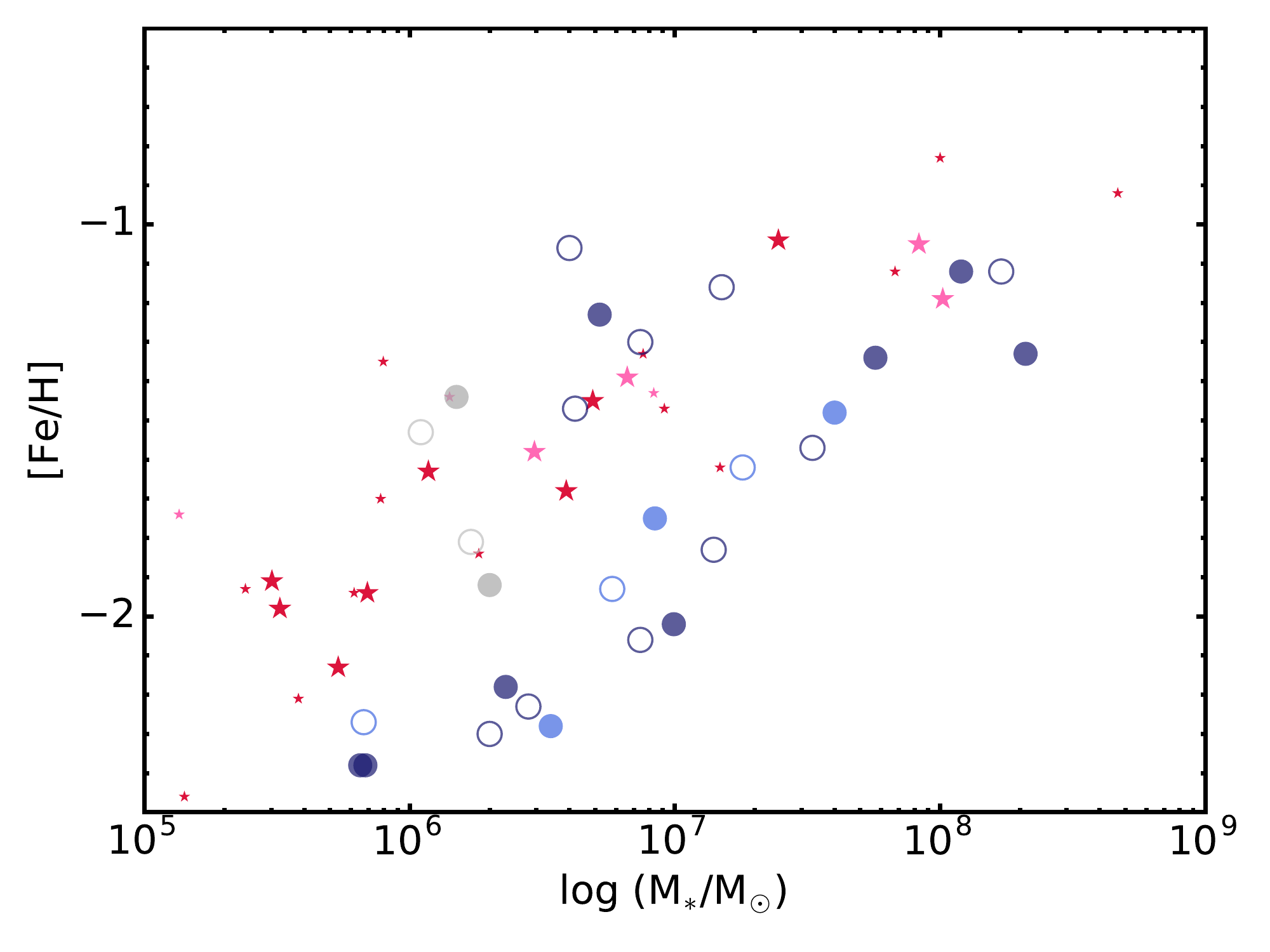}
\caption{A comparison between the stellar mass-metallicity relation (MZR) at $z$ = 0 for different definitions of the mean metallicity of simulated galaxies (\S~\ref{sec:metal_def}). All data analysed in this paper are shown, including isolated dwarfs galaxies from FIRE simulations (grey circles) (Table~\ref{tab:sims}), dwarf galaxies from the Latte simulation (blue circles) (\S~\ref{sec:latte}), and observed LG dwarf galaxies (red stars) (Table~\ref{tab:obs}). The label TMD indicates the presence of sub-grid turbulent metal diffusion (filled circles). Additional observations of LG dIrrs and  M31 dSphs are included (small red stars) \citep{Kirby2013}. 
We do not show observational errors for clarity. (\textit{Top}) The MZR for $\langle$[Fe/H]$\rangle$, defined analogously to the observational data (Eq.~\ref{eq:metal_def_ivanna}). (\textit{Bottom}) The MZR for the definition of mean metallicity used in Eq.~\ref{eq:metal_def_andrew}, based on mean metal mass fractions of entire simulated galaxies.  Note that the latter definition shows some overlap with the observations for \mstar \ > 10$^{6}$ \msun. However, the definition motivated by observational methods is systematically offset from the observed MZR by $\sim$ 0.3 dex.
\label{fig:mass_metallicity_relation}}
\end{figure}

Although metals are diffused via turbulence in a realistic ISM, this has yet to be taken into account in many galaxy evolution and formation simulations \citep{Wadsley2008}. Lagrangian codes, such as SPH and MFM, follow fluid elements of fixed mass.
Particles conserve metallicity unless injected with metals or metal loss occurs owing to SNe/stellar winds. However, SPH, or any Lagrangian methods (MFM), do not account for additional mixing that occurs via sub-grid Kelvin-Helmholtz instabilities, Rayleigh-Taylor instabilities, and turbulent eddies between gas particles. That is, without sub-grid metal diffusion, the metals assigned to a given gas particle are locked to that particle for all time. Consequently, gas particles may never become enriched, resulting in artificial noise in the MDF. Other sources of noise that may impact the appearance of the MDF  are addressed in Appendices~\ref{sec:imf} and ~\ref{sec:metal_inject}. Moreover, even enriched particles contribute to an unrealistic spread in metallicity in the absence of sub-grid mixing. To account for such unresolved mixing processes, some of the simulations include an explicit metal diffusion term between particles, following the prescription investigated by \citet{Shen2010} based on the \citet{Smagorinsky1963GeneralEquations} model,

\begin{equation}
\label{eq:metal_diff}
\begin{gathered}
\frac{\partial \textbf{M}_i}{\partial t} + \nabla \cdot (D \nabla \textbf{M}_i) = 0,\\
D = C_0 \lVert \textbf{S} \rVert_f \textbf{h}^2,
\end{gathered}
\end{equation}
where \textbf{h} is the resolution scale, and $C_0$ is proportional to Smagorinsky-Lilly constant calibrated from direct numerical simulations \citep{Su2017,Hopkins2017arXiv}. For a discussion of the coefficient calibration, see Appendix~\ref{sec:diff_coeff}. 
We adopt a value of $C_0$ $\approx$ 0.003. The symmetric traceless shear tensor is given by

\begin{equation}
\label{eq:shear_tensor}
\textbf{S} = \frac{1}{2}\left(\nabla \textbf{v} + \left(\nabla \textbf{v} \right)^T \right) - \frac{1}{3} Tr \left( \nabla \textbf{v} \right),
\end{equation}
where \textbf{v} is the associated shear velocity.

More simplistically, $ D \sim \lambda_{\textrm{eddy}} v_{\textrm{eddy}}$, where the largest unresolved eddies dominate the sub-grid diffusivity, i.e., $\lambda_{\textrm{eddy}} \sim \textbf{h}$. 
The only effect of a sub-grid prescription is to smooth variations in metallicity between fluid elements. However, since the shear tensor (Eq \ref{eq:shear_tensor}) can be artificially triggered by bulk motion such as rotation, the above model for turbulent diffusion likely over-estimates the true diffusivity. 
We further address the possibility of over-mixing and illustrate the robustness of our results with respect to the diffusion coefficient in Appendix~\ref{sec:diff_coeff_stability}.

\vspace{-0.4cm}
\section{Metallicity Distribution Functions}
\label{sec:mdf}

\subsection{Metallicity Definitions}
\label{sec:metal_def}

We analyse the stellar-mass weighted\footnote{We weight the MDFs and mean metallicities by stellar mass, though mass-weighting does not significantly impact these quantities. The FIRE simulations include standard particle splitting and merging such that no particle ever deviates from the median particle mass by more than a factor of 3. That is, 99.99\% of all star particles are within 0.2 dex of the median particle mass \citep{Hopkins2017arXiv}.} metallicity\footnote{We adopt the observational convention, where metallicity refers to stellar iron abundance ([Fe/H])}  distribution functions of the simulations to quantify the impact of metal diffusion. The abundances from the simulation are calculated from the absolute metal mass fractions per element of a star particle, 

\begin{equation}
\begin{split}
[X/Y] &= \log_{10} \left(\frac{m_Y M_X}{m_X M_Y} \right) - (\log\epsilon_{X,\odot} - \log\epsilon_{Y,\odot}),\\
\log\epsilon_X &= \log_{10} \left( N_X/ N_H \right) + 12,
\end{split}
\label{eq:metallicity}
\end{equation}
where $X$ and $Y$ represent chemical species, $m_X$ is the atomic mass for a given species, $M_X$ is the metal mass fraction, and $\epsilon_{X, \odot}$ is the abundance relative to solar \citep{AndersGrevesse1989,Sneden1992}, observationally determined from $N_{X}$, the number density of the species.

We adopt the following definition of mean metallicity, $\langle$[Fe/H]$\rangle$ (Table ~\ref{tab:sims}), motivated by observational measurements of Local Group dwarf galaxies (\citealt{Kirby2010MULTI-ELEMENT,Kirby2013}; Table ~\ref{tab:obs}),

\begin{equation}
\langle\textrm{[Fe/H]}\rangle = \frac{ \sum_i^N \textrm{[Fe/H]}_i \textrm{m}_i }{\sum_i^N \textrm{m}_i},
\label{eq:metal_def_ivanna}
\end{equation}
where [Fe/H]$_i$ is the metallicity of an individual star particle, calculated according to Eq.~\ref{eq:metallicity}, m$_i$ is the mass of the star particle, and $N$ is the total number of star particles in a given simulation.  This is in contrast to the definition of mean metallicity based on mass-averaged metal mass fractions used previously in FIRE papers \citep{Ma2016,Wetzel2016},
\begin{equation}
\begin{split}
\textrm{[Fe/H]} &= \log_{10} \left( \frac{ \bar{\textrm{f}_{\rm{Fe}}} }{ \bar{\textrm{f}_{\rm{H}}} } \right) - \log_{10}\left( \frac{ \textrm{f}_{\rm{Fe},\odot} }{ \textrm{f}_{\rm{H},\odot} } \right),\\
\bar{\textrm{f}_{\rm{Fe}}} &= \frac{ \sum_i^N \textrm{f}_{\rm{Fe},i} \textrm{m}_i }{ \sum_i^N \textrm{m}_i },
\end{split}
\label{eq:metal_def_andrew}
\end{equation}
where f$_{\rm{Fe},i}$ is the absolute iron mass fraction of a star particle. Eq.~\ref{eq:metal_def_ivanna} is the mean of the logarithm, whereas Eq.~\ref{eq:metal_def_andrew} is the logarithm of the mean. 

The Eq.~\ref{eq:metal_def_andrew} definition is appropriate for more distant galaxies,  where the stellar metallicity is determined from galaxy-integrated spectra. In this case, stellar population synthesis models are used to measure Fe and Mg, which dominate the absorption features in stellar atmospheres (e.g., \citealt{Gallazzi2005}). However, Eq.~\ref{eq:metal_def_andrew} is not consistent with mean metallicity measurements based on spectra of resolved stellar populations, i.e., LG dwarf galaxies.

In the Eq.~\ref{eq:metal_def_andrew} mean metallicity definition,  metal-rich stars are weighted  heavily, particularly for galaxies with skewed MDFs,  such that the scatter in the stellar mass-metallicity relation and the mean metallicity for such galaxies increases. As shown in the right panel of Figure~\ref{fig:mass_metallicity_relation}, this results in a $\sim$ 0.3 dex discrepancy in the FIRE stellar mass-metallicity relation between definitions of mean metallicity. Adopting the observationally motivated definition (Eq.~\ref{eq:metal_def_ivanna}) similarly results in a $\sim$ 0.3 dex offset relative to the observed mass-metallicity relation for LG dwarf galaxies, whereas the alternate definition (Eq.~\ref{eq:metal_def_andrew}) shows some overlap with observations.

The offset in the FIRE stellar mass-metallicity relation relative to observations of low-mass galaxies is likely caused by systematic uncertainties in the SNe Ia delay time distribution, and potentially the yields. The systematic offset ($\sim$ 0.03 dex) from adopting the solar abundances of, e.g., \citet{Asplund2009}, is negligible. Assuming a SNe Ia rate with prompt and delayed components \citep{Mannucci2006}, as opposed to a power-law rate (e.g., \citealt{Maoz2017arXiv}), can result in a factor $\sim$ 2 reduction in the number of SNe Ia for a fixed stellar population over 10 Gyr, given the same minimum age  for the onset of SNe Ia. Adopting a SNe Ia delay time distribution with a larger integrated number of events could therefore sufficiently increase the FIRE mean metallicity (Eq.~\ref{eq:metal_def_ivanna}) of simulated dwarf galaxies to result in better agreement with observations. Conclusively resolving this discrepancy is beyond the scope of this paper, and will be addressed in future work (Wetzel et al., in prep).

Although the offset between the FIRE stellar mass-metallicity relation and observations is $\sim$ 0.3 dex for low-mass (\mstar \ $\lesssim$ 10$^{9}$ \msun) galaxies,  it only impacts the metallicity normalization, as opposed to comparisons of the overall MDF shape, the width of the MDF, and the intrinsic scatter in [$\alpha$/Fe] vs. [Fe/H]. In what follows, [Fe/H] refers to the Eq.~\ref{eq:metal_def_ivanna} definition.  

\vspace{-0.3cm}
\subsection{Narrowing Effect of Turbulent Metal Diffusion}
\label{sec:mdf_width}

\begin{figure}
\centering
\includegraphics[width=\columnwidth]{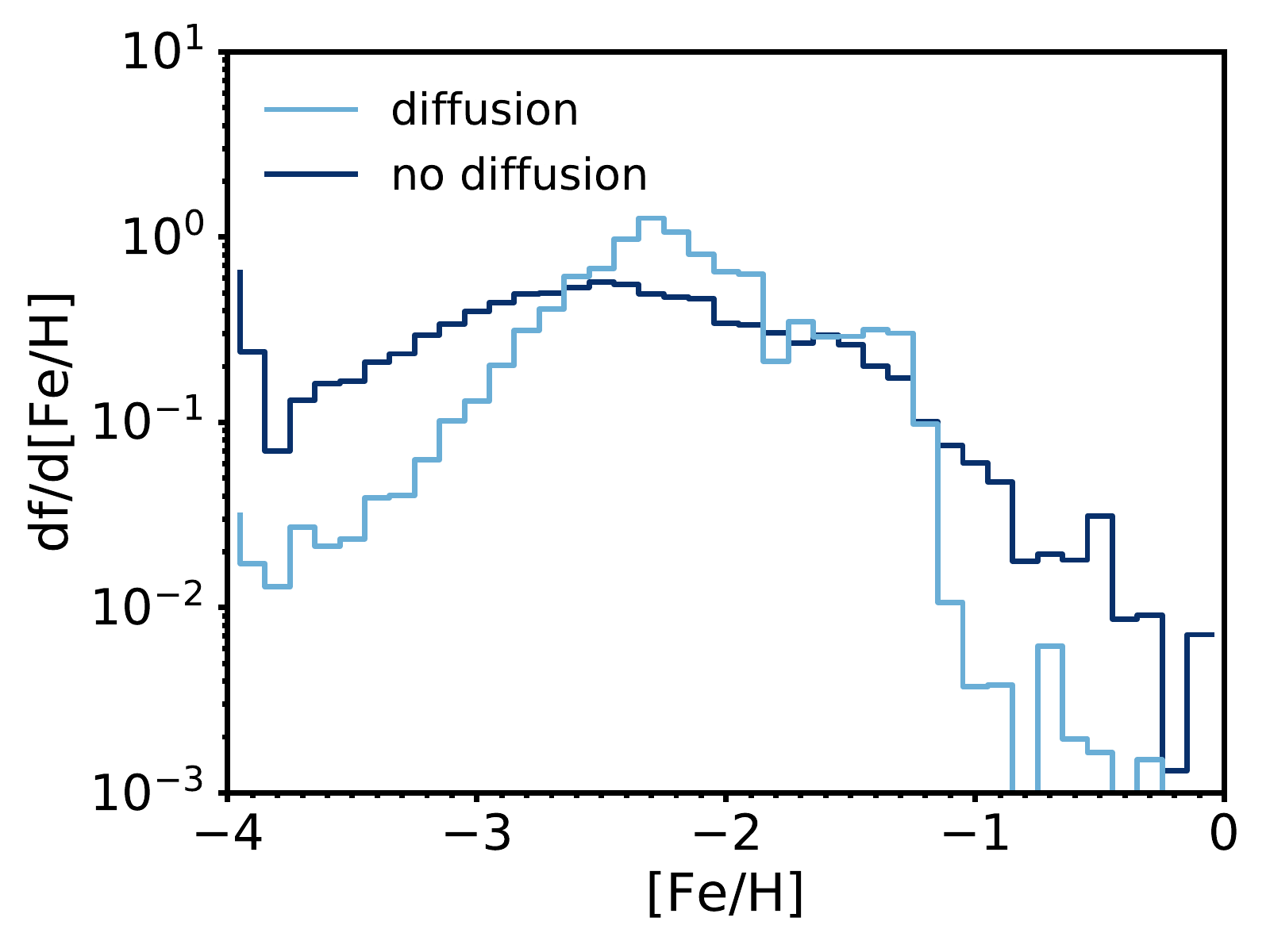}
\hspace{-0.5cm}
\caption{An example of the comparison between the stellar mass-weighted metallicity distribution functions of stars in the galaxy at $z$ = 0 with and without turbulent metal diffusion for \textbf{m10q}. The metallicity distribution functions are plotted in terms of probability density on a log scale to emphasize the behaviour at the tail of the distributions. The narrowing effect of diffusion is clear, resulting in a reduction in the width of the MDF by 0.14 dex, or a factor of 1.4 (40\%). Similar behaviour occurs in the case of \textbf{m10v}. \label{fig:mdf_comp_theory}}
\end{figure}

\begin{table*}
\caption{Properties of Local Group Dwarf Galaxies\label{tab:obs}}
\begin{threeparttable}
\begin{tabular*}{\textwidth}{l @{\extracolsep{\fill}} ccccc}
	\toprule
	Galaxy & D (kpc)\tnote{a} & $\log$(\mstar/\msun)\tnote{b} & 	$\langle$[Fe/H]$\rangle$\tnote{c} (dex) & $\sigma$\tnote{d} (dex) &
    $N_{[Fe/H]}$\tnote{e} \\ \midrule
	\multicolumn{6}{l}{\textbf{MW dSphs}} \\ \\
	Canes Venatici I & 217 $\pm$ 23 & 5.48 $\pm$ 0.09 &  $-$1.91 & 0.44 (0.39) & 151 \\
	Draco & 75 $\pm$ 5 & 5.51 $\pm$ 0.10 & $-$1.98 & 0.42 (0.35) & 333 \\
	Ursa Minor & 75 $\pm$ 3 & 5.73 $\pm$ 0.20 & $-$2.13 & 0.43 (0.34) & 670 \\
	Sextans & 85 $\pm$ 3 & 5.84 $\pm$ 0.20 & $-$1.94 & 0.47 (0.38) & 96 \\
	Leo II & 233 $\pm$ 13 & 6.07 $\pm$ 0.13 & $-$1.63 & 0.40 (0.36) & 256 \\
	Sculptor & 85 $\pm$ 4 & 6.59 $\pm$ 0.21 & $-$1.68 & 0.46 (0.44) & 365 \\
	Leo I & 253 $\pm$ 15 & 6.69 $\pm$ 0.13 & $-$1.45 & 0.32 (0.28) & 774 \\
	Fornax & 147 $\pm$ 9 & 7.39 $\pm$ 0.14 & $-$1.04 & 0.33 (0.29) & 665 \\ \\
	\multicolumn{6}{l}{\textbf{dIrrs}} \\ \\
	Leo A & 787 $\pm$ 29 & 6.47 $\pm$ 0.09 & $-$1.58 & 0.42 (0.36) & 146 \\
	Peg dIrr & 920 $\pm$ 29 & 6.82 $\pm$ 0.08 & $-$1.39 & 0.56 (0.54) & 99 \\
	NGC 6822 & 459 $\pm$ 8 & 7.92 $\pm$ 0.09 & $-$1.05 & 0.49 (0.47) & 298 \\
	IC 1613 & 758 $\pm$ 4 & 8.01 $\pm$ 0.06 & $-$1.19 & 0.37 (0.32) & 132 \\ \bottomrule
\end{tabular*}
\begin{tablenotes}
\item[a] Distance from the Milky Way (\citealt{Kirby2014} and  references therein).
\item[b] Stellar masses determined by \citet{Woo2008}, with the exception of Canes Venatici I \citep{Martin2008}.
\item[c] Error-weighted mean metallicity, determined analogously to Eq.~\ref{eq:metal_def_ivanna}. All galaxies have a standard error of the mean of 0.01 dex, except Leo A, with a standard error of the mean of 0.02 dex \citep{Kirby2013}.
\item[d] The MDF width (error-corrected in parenthesis), calculated by \citealt{Kirby2013}.
\item[e]  The number of confirmed radial velocity members with [Fe/H] $>$ $-$3 dex and measurement uncertainty $\leq$ 0.5 dex.
\end{tablenotes}
\end{threeparttable}
\end{table*}

Compared to simulations without sub-grid diffusion, we observe a narrowing of the characteristic width of the MDF when including sub-grid diffusion\footnote{We acknowledge the potential of run-to-run variations in stellar mass and SFH due to stochasticity caused by random system perturbations \citep{Su2017}. Any individual detailed feature in the MDFs could be due to stochastic fluctuations, but general MDF properties such as the reduction of the MDF width and behaviour at the tails are retained in statistical populations of simulated dwarf galaxies (\S~\ref{sec:latte}). Stochastic effects are generally small in magnitude compared to the magnitude of systematic effects we observe in the MDF and $\alpha$-element abundance ratio distributions.} (Figure~\ref{fig:mdf_comp_theory}). For \textbf{m10q}, the standard deviation of the MDF is reduced from 0.55 dex to 0.41 dex (a factor of $\sim$ 1.4), whereas for \textbf{m10v} it is reduced from 0.52 dex to 0.34 dex (a factor of $\sim$ 1.5, Table~\ref{tab:sims}).  This is in better agreement with the MDF width for a majority of  the LG dwarf galaxies  (Table  ~\ref{tab:obs}), particularly for those within the mass range spanned by the simulations.

Although this narrowing effect of the MDF  when including diffusion may initially seem counterintuitive, it is in accordance with expectations, given that individual particles are being driven toward the average metallicity as a result of sub-grid mixing.  Star particles are no longer born on the low-metallicity tail of the distribution ([Fe/H] $\lesssim$ $-$3 dex), which corresponds to extreme, improbably low metallicities, or the high-metallicity tail ([Fe/H] $\gtrsim$ $-$0.5 dex), which corresponds to metallicities that are not observed in most LG dwarf galaxies.

The Milky-Way mass FIRE simulation \textbf{m12i} (\S 5) \citep{Wetzel2016} (\mstar \ $\sim$ 6.5 $\times$ 10$^{10}$ \msun), including sub-grid metal diffusion, also exhibits a narrowing of the MDF. Considering that the effects of metal mixing microphysics on galaxy dynamics, as well as other global galaxy properties, are negligible \citep{Su2017}, our results are likely applicable to dwarf galaxy simulations within a broad mass range. 

\vspace{-0.3cm}
\subsection{Comparison to Observed MDFs}
\label{sec:mdf_comp}

Next, we investigate whether the narrowed theoretical MDFs are in better agreement with observations via comparison to those of MW satellite dwarf spheroidal (dSph) galaxies and LG isolated dwarf irregular (dIrr) galaxies (Table~\ref{tab:obs}), for which we have metallicity measurements of $\gtrsim 100$ red giants per galaxy \citep{Kirby2010MULTI-ELEMENT,Kirby2013}.\footnote{We do not anticipate any bias due to ``mass-weighting'' of red giants (\mstar \ $\sim$ 0.8 \msun) in the mean metallicity or MDF for dSphs, which have uniformly old stellar populations. However, dIrrs contain intermediate age stellar populations, with higher metallicity and longer lifetimes on the red giant branch. For this reason, the mean metallicity of dIrrs may be slightly biased toward higher metallicity, and the MDF width may also be affected \citep{Kirby2017,ManningCole2017}}

To determine the similarity between the observed and simulated MDFs, we quantify the likelihood that the observed stars could have been drawn from the simulated MDF. The log-likelihood is given by $\ln L$,

\begin{equation}
L = \prod_i^n L_i,
\end{equation}

\begin{equation}
\begin{split}
L_i = &\int \rm{d}\rm{[Fe/H]} \frac{dP}{d\rm{[Fe/H]}} \\
&\times \left[ \frac{1}{\sqrt{2 \pi} \sigma(\rm{[Fe/H]})_{i}} \exp(\frac{-(\rm{[Fe/H]} - \rm{[Fe/H]}_{i})^2}{2 \sigma(\rm{[Fe/H]})_{i}^2} \right],\\
&\frac{dP}{d\rm{[Fe/H]}} = \frac{1}{m} \sum_j^m \delta(\rm{[Fe/H]})_j
\end{split}
\label{eq:max_like}
\end{equation}
where $n$ is the number of measurements for a given observed dwarf galaxy, $i$ corresponds to an individual measurement (i.e. red giant),  $m$ is the number of star particles in a given simulation, and  $j$ corresponds to an individual star particle. $\sigma$([Fe/H]) is the observed measurement uncertainty in metallicity, and $\delta$([Fe/H]) is a delta function for a star particle with a given metallicity.

\begin{figure*}
\centering
\includegraphics[width=\columnwidth]{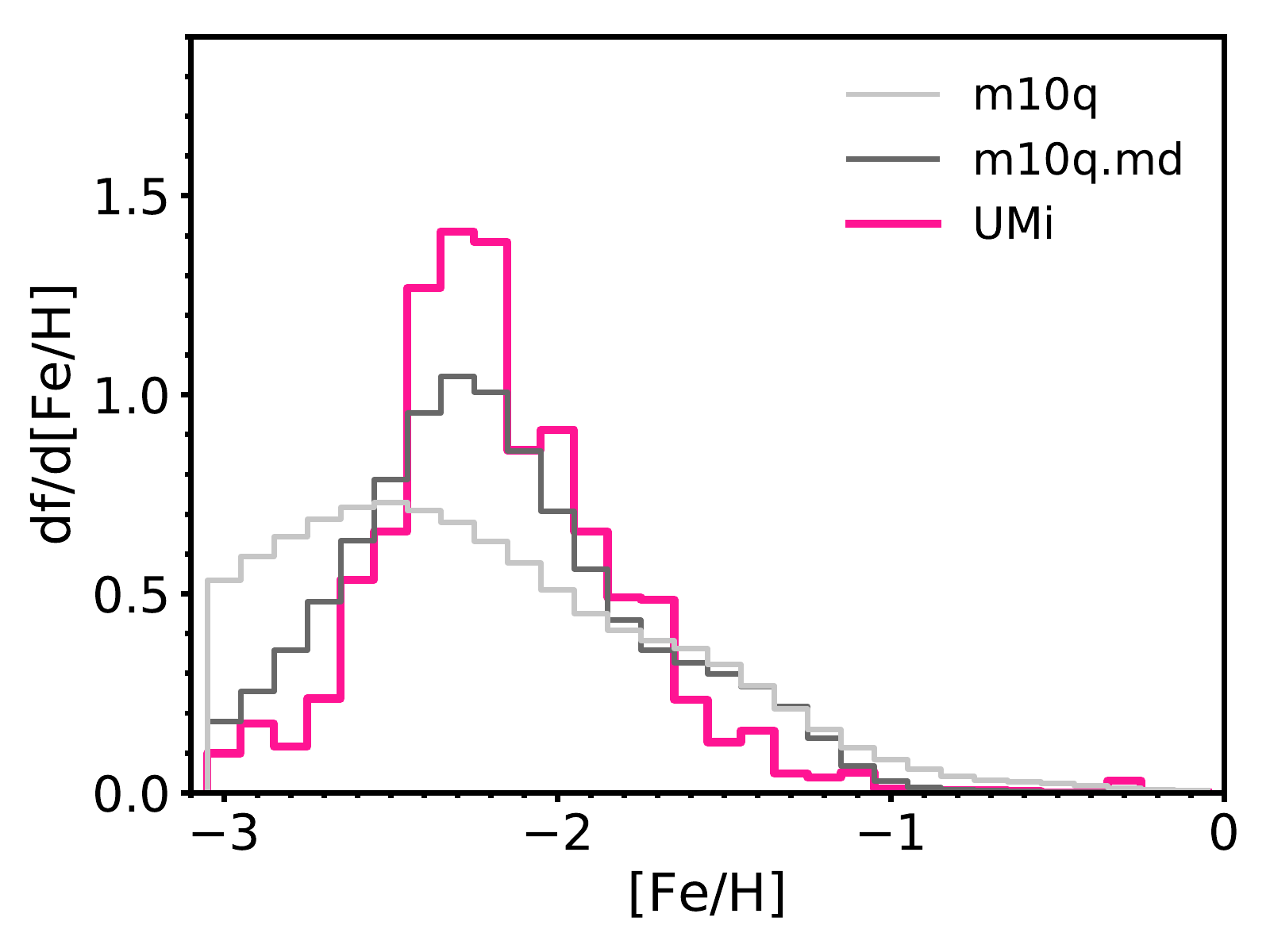}
\hspace{-0.5cm}
\includegraphics[width=\columnwidth]{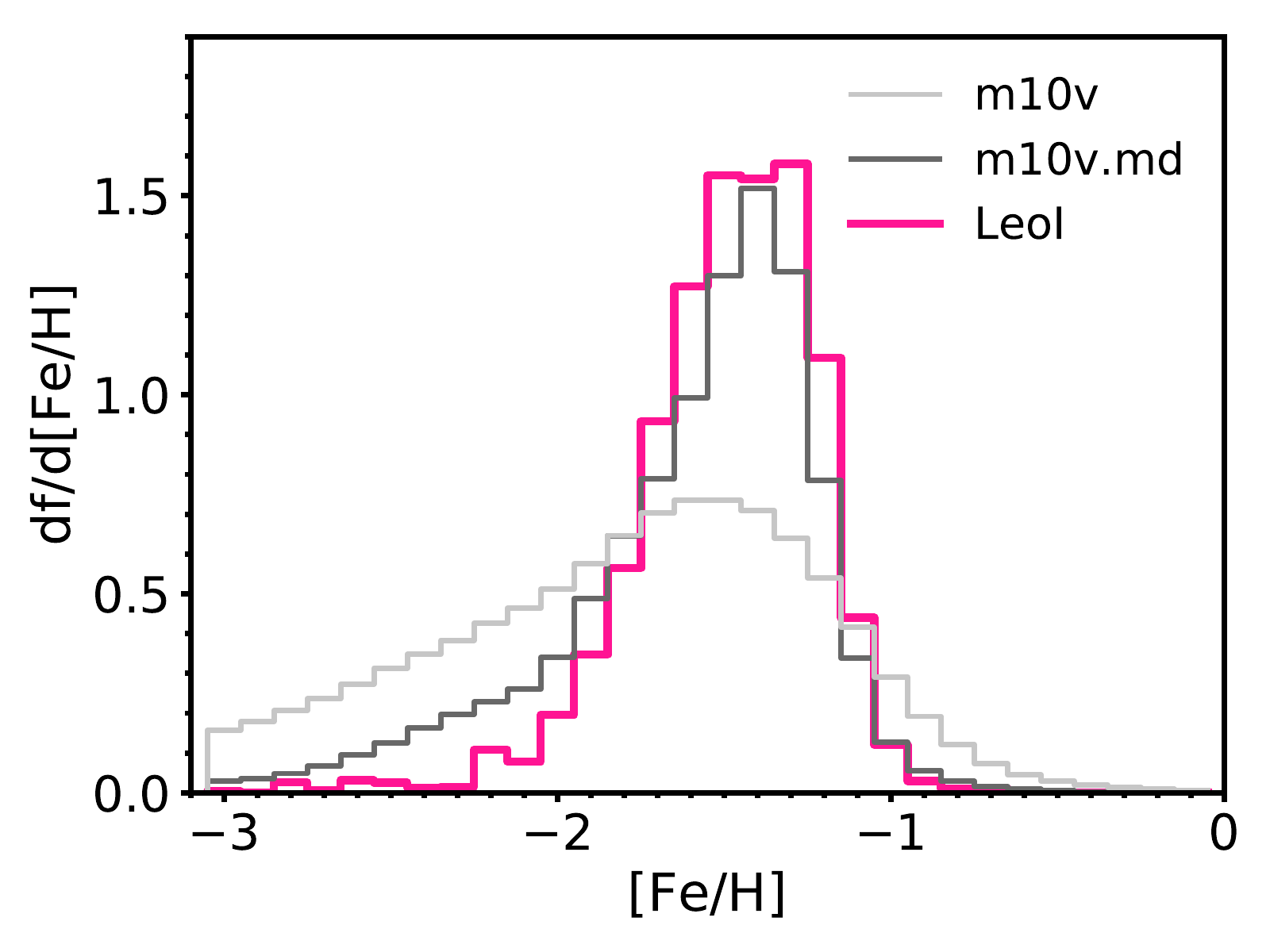}
\caption{A comparison between the MDFs for \textbf{m10q} (left) and \textbf{m10v} (right) at $z$ = 0 (\mstar \ $\sim$ 1 - 2 $\times$ 10$^{6}$ \msun) with and without TMD plotted against observational data \citep{Kirby2010MULTI-ELEMENT} of MW dSph satellite galaxies (Ursa Minor, Leo I: \mstar \ $\sim$ 5.4 $\times$ 10$^{5}$, 4.9 $\times$ 10$^{6}$ \msun \ respectively; \citealt{Woo2008}). The two observed dSph MDFs exhibit the highest likelihood of being drawn from these two simulations. The theoretical MDFs are weighted by stellar mass\usebox2 and have been smoothed to reproduce the effects of observational uncertainty.  By comparison with the MDFs without sub-grid metal mixing, TMD appears to better reproduce the narrowness, and overall shape, of the dwarf galaxy MDFs. \label{fig:mdf_comp_obs}}
\end{figure*}

We approximate the probability distribution for the theoretical metallicity as a sum of a delta functions with no associated error. When computing the likelihood between each pair of simulated and observed galaxies, we exclude measurements with $\sigma$[Fe/H] $>$ 0.5 dex from the observational data and [Fe/H] $<$ $-$3 dex from both the observational and simulated data sets. At such low metallicity, it is more likely for the metal content of star particles to be dominated by a single SN event in the simulations. It is also possible for the star particles to be dominated by relics of Population III stars, owing to the initialization of particles at [Fe/H] = $-$4 in the absence of explicit modelling of the transition to Population II stars. The fraction of stars in the relevant stellar mass range for dwarf galaxies (10$^{6}$ \msun \ $\lesssim$ \mstar \ $\lesssim$ 10$^{8}$ \msun) with [Fe/H] $<$ $-$3 is $\lesssim$ 4.5\% for the high-resolution simulations, such that it impacts the detailed shape of the low-metallicity tail of the distribution, as opposed to the MDF width. Thus, we exclude the potentially unphysical, extremely metal-poor stars present in the simulations that are not seen in observational data.  

Figure~\ref{fig:mdf_comp_obs} shows the results of the likelihood estimation method of comparison, where the theoretical MDFs are mass-weighted and smoothed to reproduce the effects of observational uncertainty. As an example, we consider the highest likelihood case for each simulated dwarf galaxy, where we compared to 12 LG dwarf galaxies with a variety of stellar masses and SFHs. We emphasize that the effects of including TMD, in relation to observations as discussed below, are generalizable to average populations of simulated dwarf galaxies (\S~\ref{sec:latte}).

MW dSphs Ursa Minor (UMi) and Leo I (\mstar \ $\sim$ 5.4 $\times$ 10$^5$ \msun \ and 4.9 $\times$ 10$^6$ \msun \ respectively; \citealt{Woo2008}) have the most statistically similar MDFs, according to Eq.~\ref{eq:max_like}, as compared to both simulations \textbf{m10q/m10q.md} and \textbf{m10v/m10v.md} respectively.\footnote{In the case of Ursa Minor and \textbf{m10q}, we observe a 0.6 dex discrepancy between the simulated and observed galaxy stellar masses for similar mean metallicity. This results in an offset in the FIRE stellar mass-metallicity relation relative to observations. This offset is likely a consequence of our choice of Type Ia SNe delay time distribution, as discussed in \S~\ref{sec:metal_def}. We emphasize that the offset in the FIRE stellar mass-metallicity relation at low stellar masses does not alter any of our conclusions regarding the MDF shape, or scatter in [$\alpha$/Fe] at fixed [Fe/H] (\S~\ref{sec:int_scat_sim}).} The \textbf{m10q.md}/UMi and \textbf{m10v.md}/Leo I pairs also have the highest average\footnote{The average log-likelihood is defined as $\hat{L} = \frac{1}{n} \ln L$, where $n$ is the number of measurements for a given observed dwarf galaxy. $\hat{L}$ estimates the expected log-likelihood of a single observation, given the ``model'' simulation. This estimator enables comparisons between different sample sizes, i.e., different LG dwarf galaxies, for a given simulation.} likelihoods compared to all \textbf{m10q}/LG dwarf galaxy and \textbf{m10v}/LG dwarf galaxy pairs. The \textbf{m10q}/UMi pair has the highest probability of all LG dwarf galaxies of being drawn from that particular simulation. For \textbf{m10v}, comparisons to Leo II and Leo I yield similar likelihoods, such that one simulation is not strongly favoured over the other. In the following discussion, we compare \textbf{m10v} to Leo I given that it is strongly favoured by \textbf{m10v.md}.

Each highest-likelihood pair of simulated and observed dwarf galaxies have similarly shaped SFHs \citep{Weisz2014}, dominated by either an early burst of star formation (\textbf{m10q.md}/UMi) or rising late time star formation (\textbf{m10v.md}/Leo I). For \textbf{m10q} vs.  UMi, $\ln L$ = $-$437, whereas for \textbf{m10q.md} vs. UMi,  $\ln L$ =  $-$343. For \textbf{m10v} vs. Leo I, $\ln L$ =  $-$425, whereas for \textbf{m10v.md} vs. Leo I, $\ln L$ = $-$129. It is clear both from the increase in the values of $\ln L$ between cases with and without TMD and Figure~\ref{fig:mdf_comp_obs} that TMD  improves the ability of the simulations to match observations in terms of MDF shape. In general, the observed MDFs are narrower than the simulated MDFs without TMD (Table~\ref{tab:sims},~\ref{tab:obs}). The ability of TMD to reproduce this effect results in the increased likelihood.

With the introduction of metal diffusion, it becomes possible to construct simulated and observed MDFs that are nearly indistinguishable. For both \textbf{m10q.md} and \textbf{m10v.md}, however, the simulations have a larger population of stars at low-metallicity as compared to observations.  The lack of stars at the low-metallicity tail of the observed MDF may be caused by selection effects. Observational bias, which does not significantly affect the mean metallicity,  may result in the preferential exclusion of rare, extremely metal-poor stars that tend to inhabit galaxy outskirts. These stars also may have been tidally stripped, now absent from satellite dwarf galaxy stellar populations. However, this only impacts the detailed shape of the metal-poor portion of the MDF  \citep{Kirby2013}.

Nonetheless, the tails of the distribution are significantly reduced compared to the case without sub-grid metal mixing (Figure~\ref{fig:mdf_comp_obs}). The mean metallicities of the distributions approximately coincide, although we note the offset in the normalization of the mean metallicity (\S~\ref{sec:metal_def}).  Most significantly, the shapes of the distributions in terms of skewness and kurtosis are consistent with TMD runs, but inconsistent with non-TMD runs. This indicates that metal diffusion may be necessary to bring theoretical predictions into agreement with observations in realistic simulations of the formation and evolution of low-mass galaxies.

Although we have thus far considered only field galaxy simulations, they most resemble the classical dSphs in the sample, as opposed to LG dIrrs such as NGC 6822 and  IC 1613, or even more massive dSphs such as Fornax.  This is due to the limited mass sampling of our isolated dwarf galaxy simulations, which have masses in the range corresponding to the observed classical dSphs. The likelihood estimation method described above is, to first order, sensitive to mean metallicity, which is dictated by the stellar mass of the galaxy.  By including the isolated dwarf galaxies that form in the zoom-in region well beyond the MW-mass host halo in the Latte simulation (\S~\ref{sec:latte}), we expand the current isolated galaxy simulation suite with metal diffusion to  contain more massive, metal-rich dwarf galaxies ([Fe/H] $\gtrsim$ 1.4) that may provide better analogues to more massive LG dwarf galaxies.

Overall,  it appears that TMD results in a better match to observations. However, it is not immediately clear if  metal diffusion is the only process that can narrow the MDF sufficiently to match observed galaxies. Alternatively, environmental effects, such as ram-pressure stripping \citep{LinFaber1983,Marcolini2003}, may have similar impacts on the appearance of the MDF as sub-grid metal diffusion. We explore this possibility in \S~\ref{sec:latte_sat_iso} by analysing simulated dwarf galaxies from the Latte simulation \citep{Wetzel2016}.

\vspace{-0.4cm}
\section{Alpha Element Distributions}
\label{sec:abnd}

\subsection{Narrowing of the Alpha-Element Abundance Ratio Distributions}
\label{sec:int_scat_sim}

\begin{figure}
\includegraphics[width=\columnwidth]
{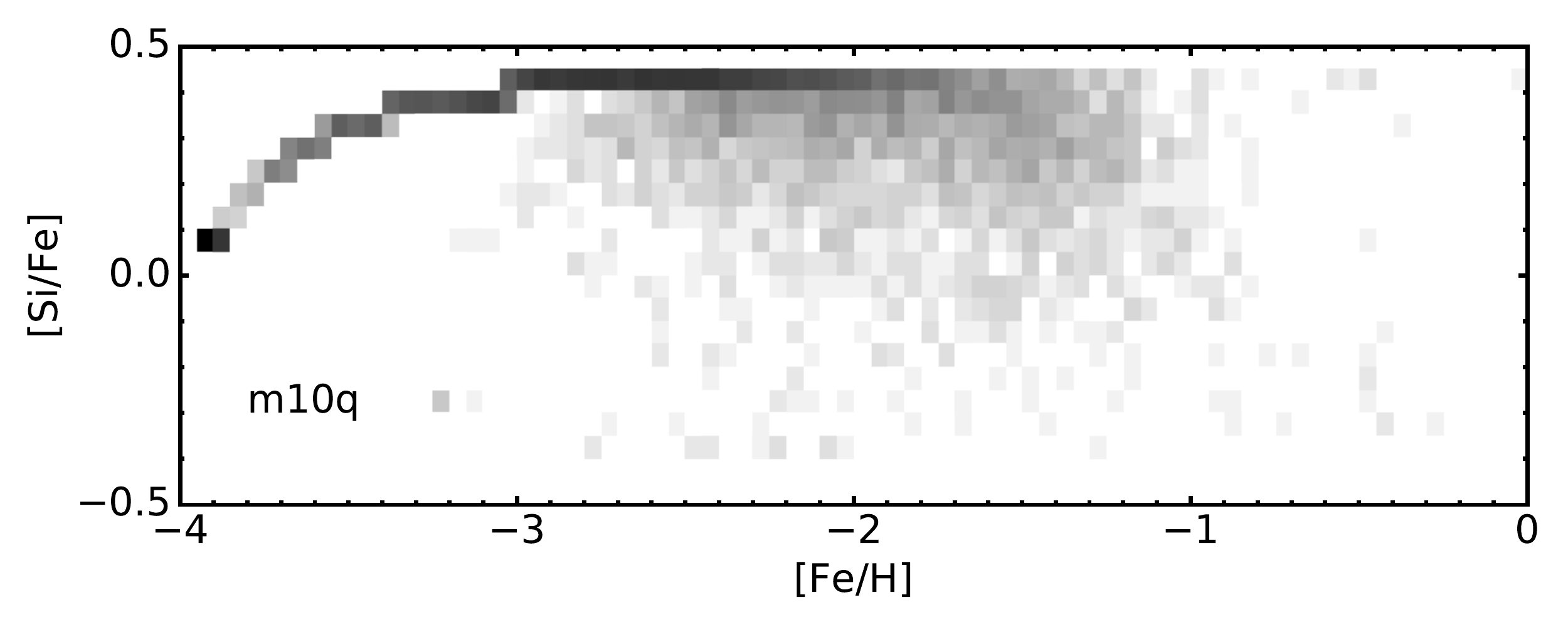}
\hspace{-0.5cm}
\includegraphics[width=\columnwidth]
{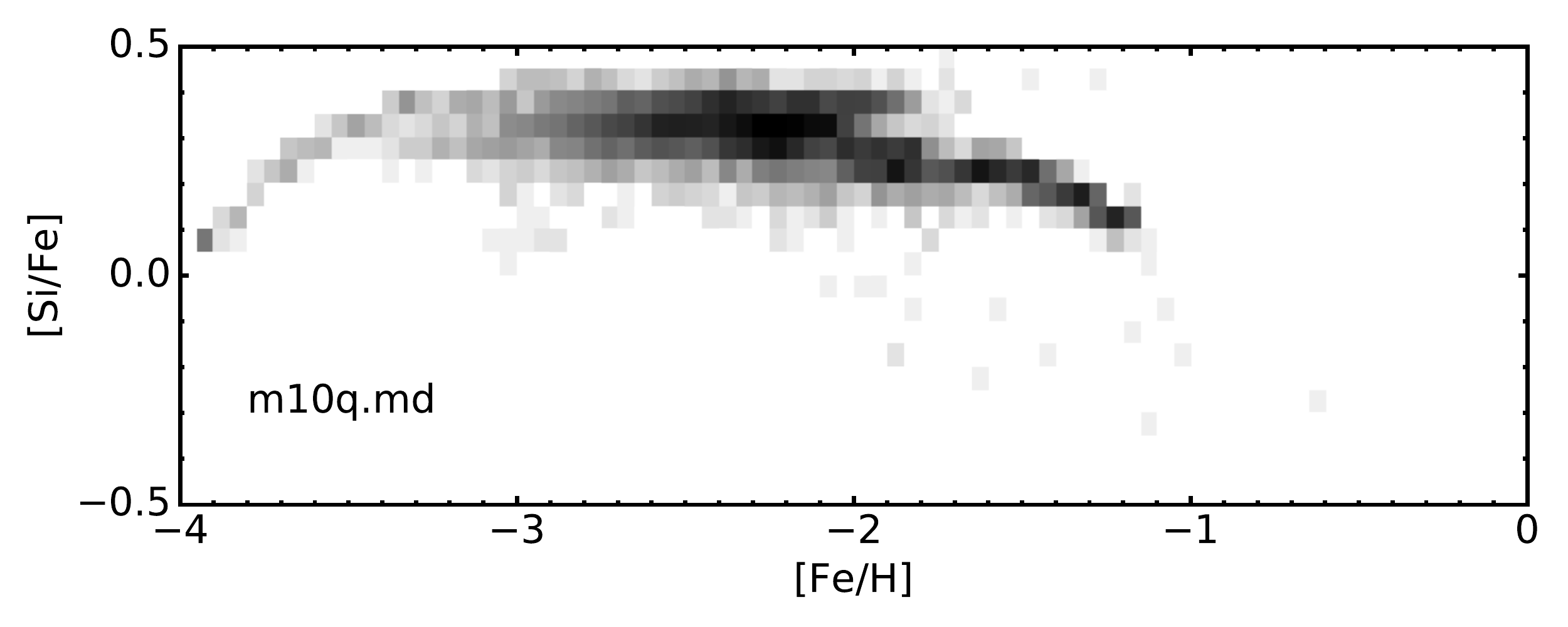}
\caption{Silicon abundance versus iron abundance for \textbf{m10q} at $z$ = 0 excluding (\textit{top}) and including (\textit{bottom}) sub-grid turbulent metal diffusion. The simulated data is represented as the relative number of star particles in 0.05 dex bins, with the darkest pixels corresponding to $\gtrsim$ 100 particles. 
The reduction in scatter of [Si/Fe] as a function of [Fe/H] is clear (0.14 dex to 0.06 dex with diffusion).  
Note that the upper (lower) envelope in the case without diffusion corresponds to enrichment by only Type II (Type Ia) SNe. The inclusion of diffusion significantly reduces the envelope for Type II SNe, whereas it vanishes completely for Type Ia SNe. 
The abundances show less dispersion at a given metallicity with diffusion turned on. This is consistent with observations, which we show in \S~\ref{sec:int_scat_obs} to have near-zero intrinsic scatter in [Si/Fe] at fixed [Fe/H]. 
\label{fig:abnd_comp}}
\end{figure}

Here, we represent the scatter in [$\alpha$/Fe] using [Si/Fe] as a proxy, owing to the lack of a theoretical analogue to measurements of $\alpha$-enhancement (\S~\ref{sec:int_scat_obs}). Figure~\ref{fig:abnd_comp} illustrates that, for the simulations, we observe the same reduction in scatter as in the MDF width for the $\alpha$-element abundance ratio distributions. There is also an apparent reduction in the envelopes corresponding to enrichment events of a single type. The envelopes originate from the  initialization of the star particles at [Fe/H] = $-$4 and [$\alpha$/Fe] = 0, where enrichment by only Type II (Type Ia) SNe results in the upper (lower) envelope. The faint lower envelope disappears entirely with the inclusion of TMD.  Sub-grid metal mixing reduces the probability that any given progenitor gas particle, where the star particle inherits the gas particles' metallicity, will only contain elements yielded from a single type of enrichment event.

In an analogous fashion to the MDF widths, we quantify the reduction in dispersion in the $\alpha$-element abundance ratio distributions caused by sub-grid turbulent metal diffusion. The intrinsic scatter in [$\alpha$/Fe] as a function of [Fe/H] has physical implications for metal mixing and chemical evolution. [Fe/H] correlates with stellar age, and so can be used as a proxy for time. Thus, a quantification of the intrinsic scatter in $\alpha$-enhancement contains information about the homogeneity of the ISM on typical galaxy evolution timescales. 

\begin{figure}
\includegraphics[width=\columnwidth]
{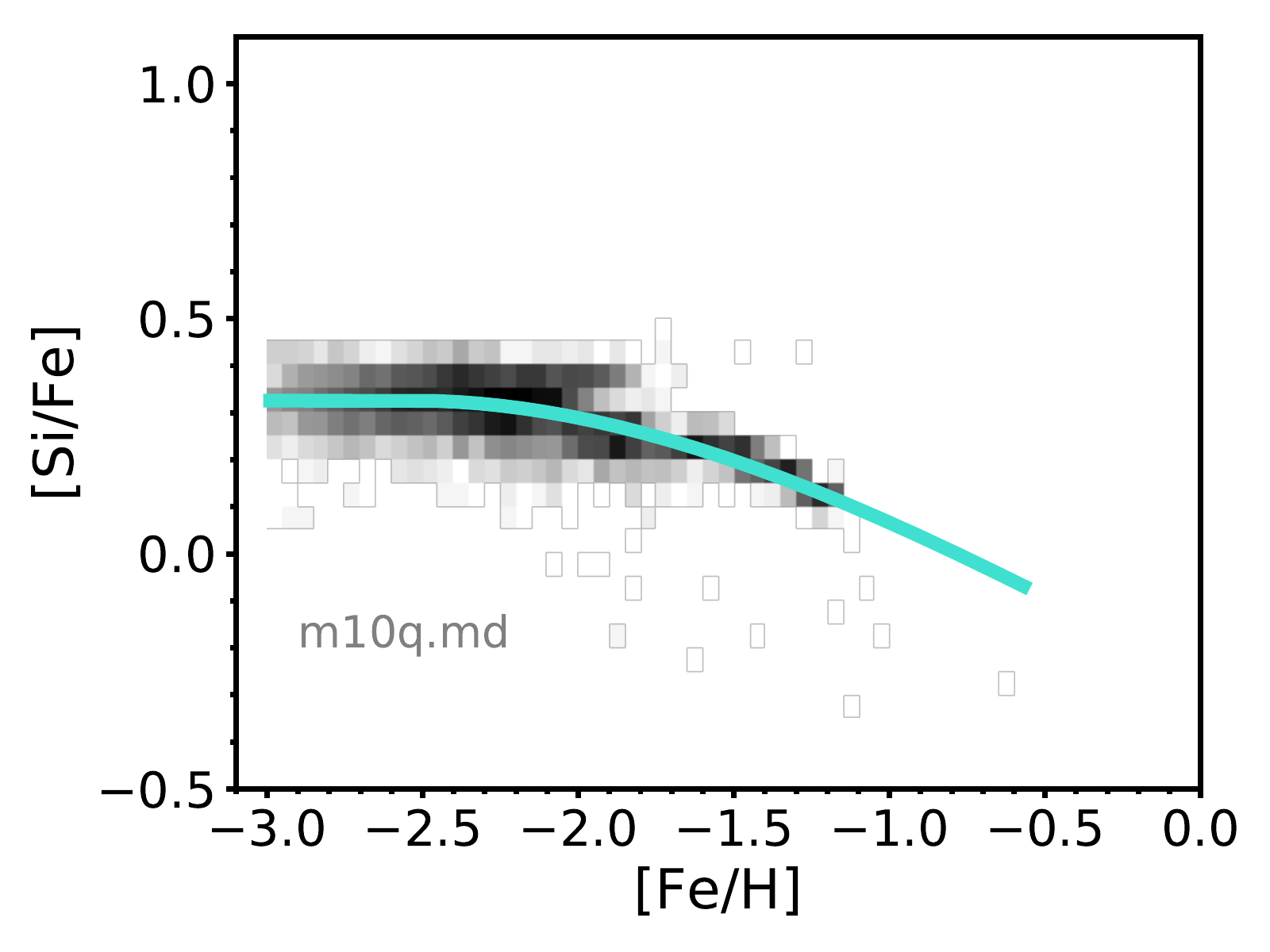}
\includegraphics[width=\columnwidth]
{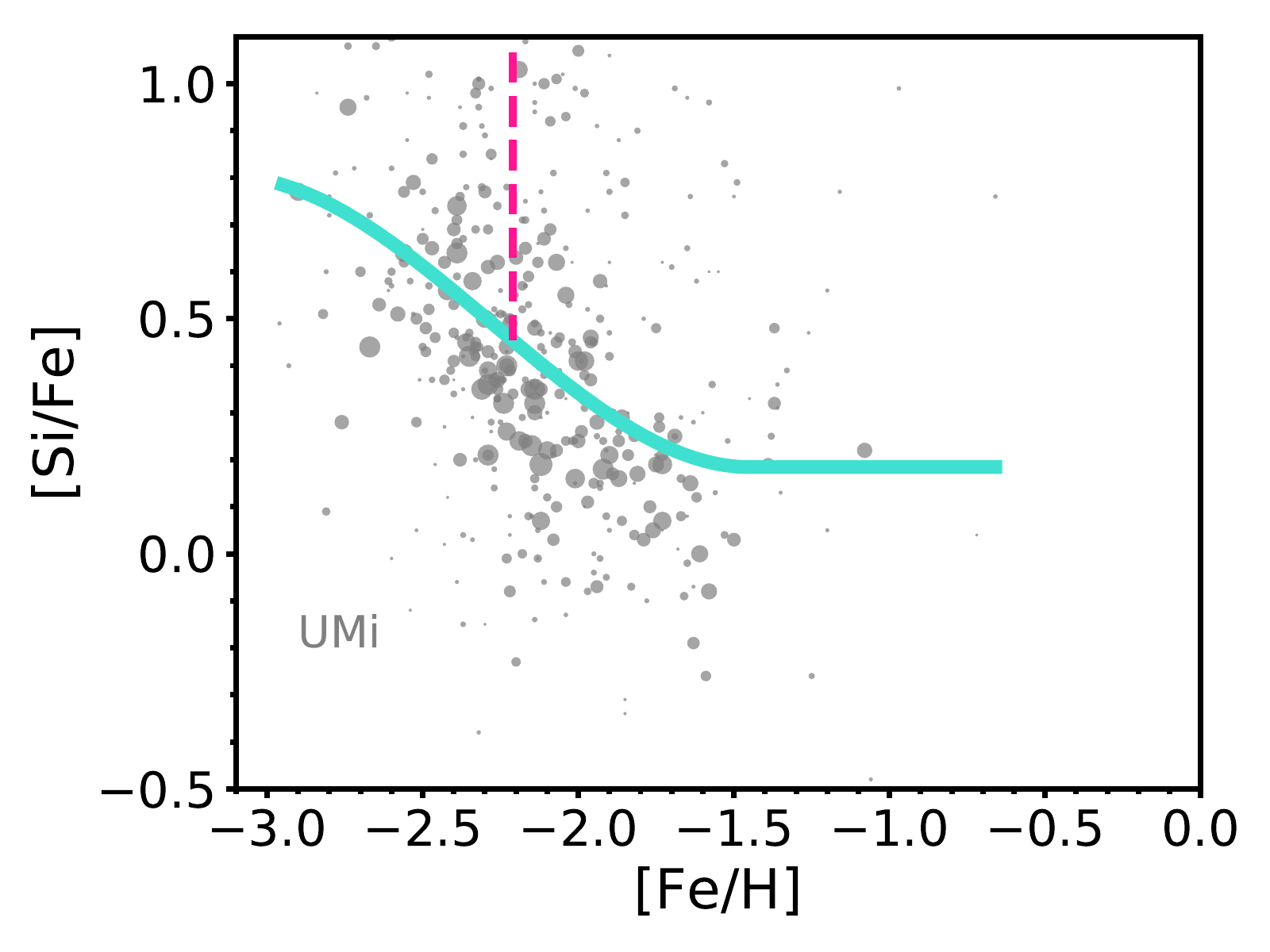}
\hspace{-0.5cm}
\caption{An illustration of the intrinsic scatter calculation for \textbf{m10q.md} (\textit{top}) at $z$ = 0 and Ursa Minor (\textit{bottom}). The simulated data is colour-coded according to the relative number of star particles in 0.05 dex bins. The size of the data points for Ursa Minor is proportional to the inverse-squared measurement uncertainty in [Si/Fe]. The turquoise curve is the best-fit cubic spline, error-weighted for the observational data, assuming that [Si/Fe] either remains constant or monotonically decreases with [Fe/H]. The dotted magenta line is the difference between a data point and the curve approximating $\langle$[Si/Fe]$\rangle$, where this difference is used to determine $\sigma$ (Eq.~\ref{eq:sigma}).}
\label{fig:int_scatter}
\end{figure}

We analyse the intrinsic scatter, $\Sigma$, in the $\alpha$-element abundance ratios as a function of metallicity for the simulations with and without sub-grid metal diffusion.  We calculate $\Sigma$ by fitting a cubic spline (error-weighted for observations) in the range $-$3 $<$ [Fe/H] $<$ $-$0.5 to the abundance ratios as a function of metallicity for both the simulated and observed LG dwarf galaxies.  We assume that [Si/Fe] either remains constant or monotonically decreases with [Fe/H]. To calculate the intrinsic scatter at fixed [Fe/H], we first determine the distance in [Si/Fe] from the curve for each data point (Figure~\ref{fig:int_scatter}).  The distribution of distances provides a standard deviation, which is the intrinsic scatter for the simulated data.  Table~\ref{tab:int_scatter} contains the results for the simulations.

The scatter for \textbf{m10q} reduces from $\Sigma$ = 0.14 dex to 0.06 dex with sub-grid diffusion, whereas for \textbf{m10v}, which is dominated by late-time star formation, $\Sigma$ = 0.13 to 0.03 dex. This corresponds to a reduction in the width by a factor of $\sim$ 1.2 and 1.3 respectively, which is comparable to the narrowing factor of the MDF width (\S~\ref{sec:mdf_width}). The near zero value of $\Sigma$ in the case of TMD implies that metals in the cool gas of the ISM are well-mixed at any given [Fe/H], or time in the galaxy's evolutionary history. We show this explicitly for the simulations in \S~\ref{sec:mdf_width_form_time}.

\vspace{-0.3cm}
\subsection{The Intrinsic Scatter at Fixed Time}
\label{sec:mdf_width_form_time}

To identify the origin of the low intrinsic scatter in $\alpha$-elements at $z$ = 0 in the simulations, we calculate the scatter at a fixed time in a galaxy's history.  Using the formation times of star particles in the galaxy (within $r_{90}$; Table~\ref{tab:sims}) at $z = 0$, we determine the dispersion in [Si/Fe], [Si/H], and [Fe/H] of star particles formed in 100 Myr time bins, a timescale comparable to the typical dynamical time of dwarf galaxies.

We base our analysis on $z$ = 0 star particles partly because it is analogous to observational methods. Additionally, we expect a negligible contribution from mergers to the $z = 0$ stellar population in dwarf galaxies. As a check, we followed the evolution of simulated dwarf galaxy progenitors to high-redshift. Using a definition of in-situ formation within 10 kpc of the galactic centre, we found that $\gtrsim$ 98\% of star particles present in the galaxy at $z = 0$ formed in-situ. \citet{AnglesAlcazar2017} and Fitts et al., in prep. similarly found in a detailed study that ex-situ star formation contributes negligibly to the stellar mass growth of isolated dwarf galaxies. In principle, all star particles from these mergers could form a distinct [Si/Fe] vs. [Fe/H] track. In this way, all star particles brought in by mergers could contribute to, or even dominate the outliers of, the scatter in abundances at a given age. Despite this, we assume that a significant majority of stars present in simulated dwarf galaxies at $z = 0$ formed in-situ, with a negligible contribution from mergers.

\begin{figure}
\centering
\includegraphics[width=\columnwidth]{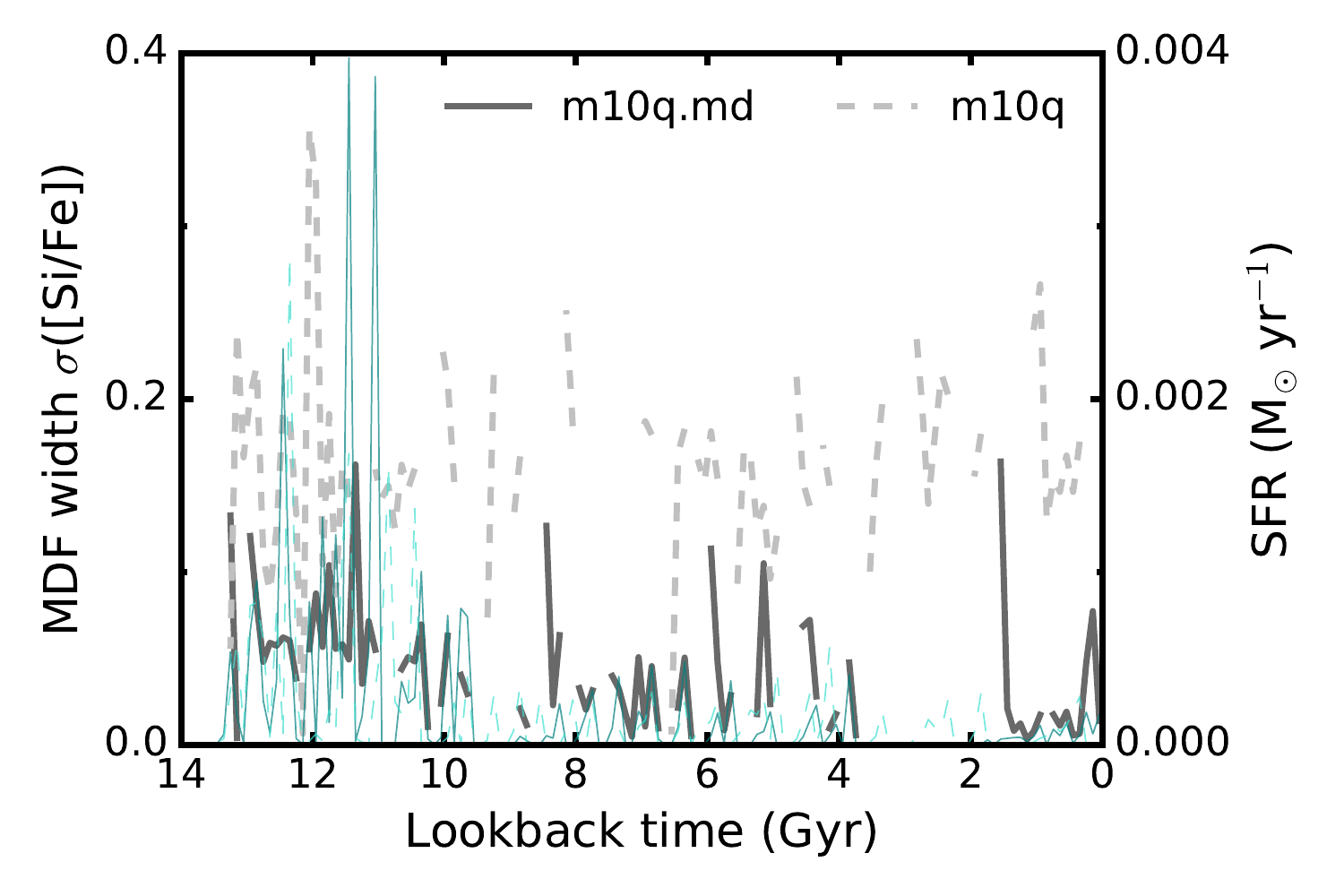}
\hspace{-0.5cm}
\includegraphics[width=\columnwidth]{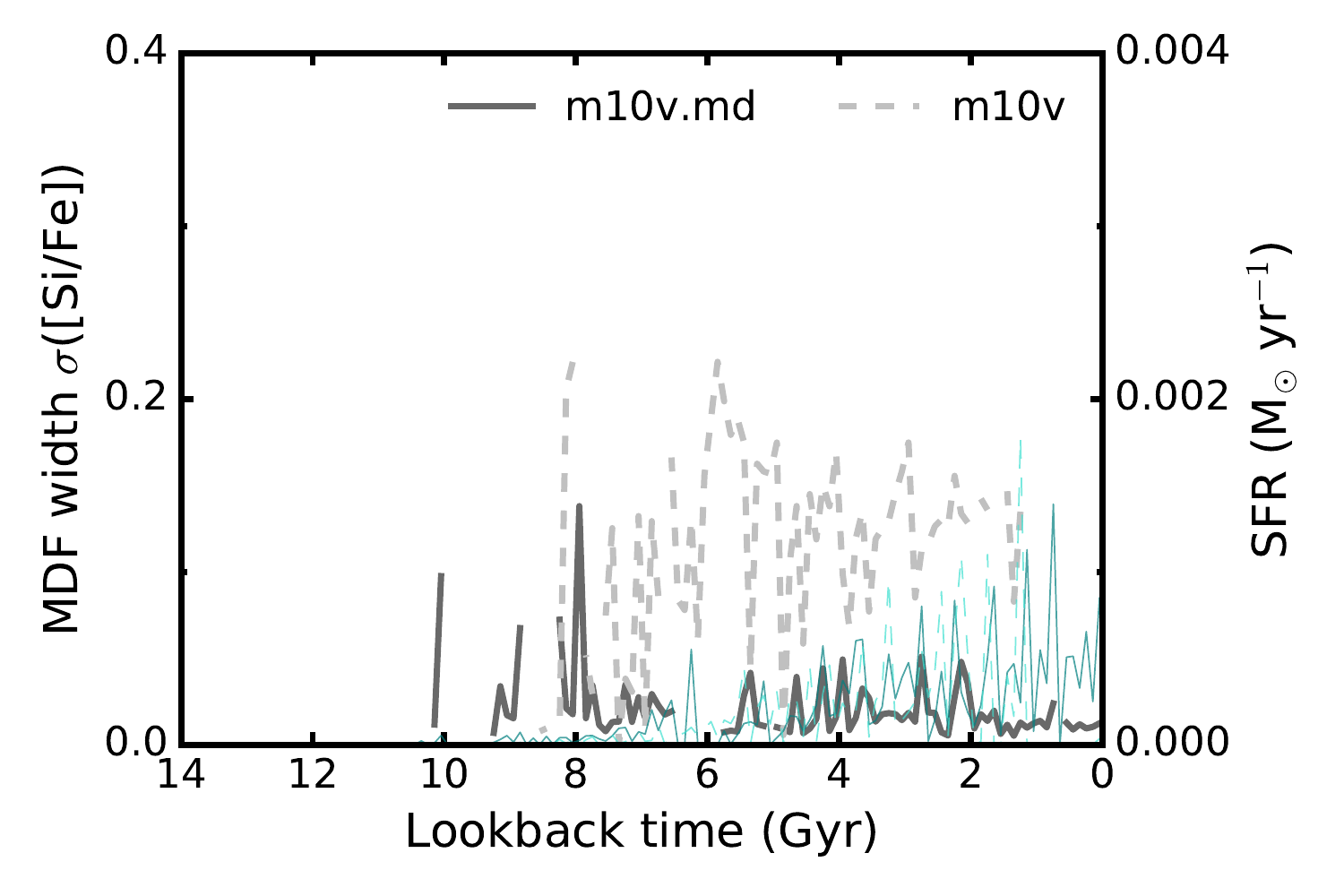}
\caption{The standard deviation of the distribution of [Si/Fe], $\sigma$([Si/Fe]) (grey), for star particles formed in 100 Myr time windows versus lookback time in Gyr. Star formation rates (turquoise) are overplotted for comparison. Both \textbf{m10q} (\textit{top}) and \textbf{m10v} (\textit{bottom}) are shown for cases with (solid lines) and without (dashed lines) sub-grid metal diffusion. The typical scatter in [Si/Fe] at a fixed time is $\sim$ 0.05 - 0.1 dex, where most of the scatter in [Si/Fe] at a fixed time is caused during short bursts of star formation. This implies that a majority of the $z$ = 0 scatter in the [Si/Fe] distribution (0.06 and 0.03 for \textbf{m10q.md} and \textbf{m10v.md} respectively) is due to time evolution of the quantity, as opposed to significant scatter in the ISM at a fixed time.
\label{fig:mdf_width_form_time}}
\end{figure}

Figure~\ref{fig:mdf_width_form_time} illustrates the scatter in [Si/Fe], presented as a standard deviation, $\sigma$([Si/Fe]), with respect to lookback time for FIRE isolated dwarfs with and without sub-grid metal diffusion. We describe the scatter in terms of [Si/Fe] for consistency with the intrinsic scatter calculation (\S~\ref{sec:int_scat_sim}). We overplot the star formation rates, where \textbf{m10q/m10q.md} have SFHs dominated by early bursts, whereas late-time star formation dominates in \textbf{m10v/m10v.md}. The scatter in [Si/Fe] as a function of age clearly reduces with the inclusion of TMD. At any given time, the typical scatter in all quantities ([Si/Fe], as well as [Si/H] and [Fe/H]) ranges from $\sim$ 0.05 - 0.1 dex. The trend between scatter and age is characterized by mostly near-zero scatter, punctuated by periods of a relatively high rate of star formation that cause the ISM to become inhomogeneous and increase the scatter. We conclude that most of the scatter in [Si/Fe] (and [Si/H] and [Fe/H] by extension) at fixed time results from starbursts. However, the persistent near-zero scatter in the abundances of newly formed stars across cosmic time implies that most of the scatter present in stellar abundance distributions at $z = 0$ is caused by time evolution, as opposed to significant scatter in the abundances of newly formed stars at any fixed time.

The fact that [Si/Fe] shows small scatter at any given time, where the naive expectation is that it should depend on both [Si/H] and [Fe/H], implies that Si enrichment correlates strongly with Fe enrichment. This is likely a result of the Type Ia and II yields \citep{Iwamoto1999,Nomoto2006} and SN rates \citep{Mannucci2006,Leitherer1999} employed in FIRE. Type Ia SNe produce a Fe yield an order of magnitude larger than that of Type II SNe, whereas the total Si yield integrated over several Gyr differs only by a factor of 2 between the SNe II and Ia. The cumulative number of core-collapse events per star particle at ages of $\sim$ 100 Myr - 1 Gyr (corresponding to the delay time of Ias in FIRE; \citealt{Hopkins2017arXiv}) is 10 - 20 times larger than the integral number of SNe Ia explosions. As a consequence, the total amount of Fe produced is comparable between SNe II and Ia channels, whereas core-collapse singly dominates Si production. Thus the correlation between Si and Fe enrichment.

We thus conclude that (1) the most likely value of the intrinsic scatter in the abundances of stars forming at any given time across a simulated galaxy's evolutionary history is near-zero, (2) deviations from near-zero scatter at a given time result from starbursts, and (3) Si enrichment correlates strongly with Fe enrichment in FIRE. The ISM is well-mixed at all times in the simulations when taking into account TMD, excepting brief starburst periods. If the intrinsic scatter in [Si/Fe] vs. [Fe/H], where [Fe/H] approximates age, is also small for observations of LG dwarf galaxies, then observed dwarf galaxies have a nearly homogeneous ISM at a given time.

\vspace{-0.3cm}
\subsection{Observational Intrinsic Scatter}
\label{sec:int_scat_obs}

To determine if TMD produces results consistent with observations, we calculate the observational intrinsic (error-corrected) scatter. 
Despite having measured [$\alpha$/Fe] directly using spectral synthesis, simultaneously fitting all $\alpha$-element lines present in the spectra, we choose to represent the intrinsic scatter in the observations by a single $\alpha$-element. This is a consequence of the difficulty in interpreting a fit based on multiple chemical species and constructing a theoretical abundance ratio counterpart. [Mg/Fe] tends to have large uncertainties, resulting in fewer data points and a less precise determination of the scatter. Although [Ca/Fe] can be measured with higher precision, the theoretical [Ca/Fe] abundances do not agree with simulations for very low metallicity (\S~\ref{sec:abnd_comp}). Thus, we adopt [Si/Fe] as a proxy for determining the intrinsic dispersion in $\alpha$-elements as a function of [Fe/H].

In the following calculation, we only consider the measurement uncertainty in [Si/Fe], as opposed to simultaneously taking into account the uncertainty in [Fe/H], $\delta$[Fe/H]. In comparison to the range of the data and the scales over which the slope of the relationship between [Si/Fe] and [Fe/H] varies,  $\delta$[Fe/H] is insignificant. The typical value of d[Si/Fe]/d[Fe/H] varies between $\sim$ 0.1 - 0.2 and $\sim$ 0.2 - 0.8  for the simulations and observations respectively, which corresponds to a variation in [Si/Fe] of $\sim$ 0.01 - 0.02 and $\sim$ 0.02 - 0.08 dex for a typical $\delta$[Fe/H] $\sim$ 0.1 dex.  This is small compared to the variation in [Si/Fe] given a typical $\delta$[Si/Fe] of $\sim$ 0.2 dex. We therefore conclude that the impact on our calculation of the intrinsic scatter owing to measurement uncertainty in [Fe/H] is negligible, such that  the uncertainty in [Fe/H] can be reasonably neglected.

For the observational case, we first calculate $\sigma^2$, the variance of the distribution of distances for each galaxy (Figure~\ref{fig:int_scatter}), normalizing to the measurement uncertainty, i.e.,

\begin{equation}
\sigma^2 =  \textrm{var} \left[ \frac{\rm{[Si/Fe]}_i - \textrm{spline}(\rm{[Fe/H]}_i)}{ \left( \delta\rm{[Si/Fe]}_i^2 + \Sigma^2 \right)^{1/2}} \right],
\label{eq:sigma}
\end{equation}
where $i$ is the index  for a given red giant with measurements of  both [Fe/H] and [Si/Fe], $\delta$[Si/Fe] is the measurement uncertainty in [Si/Fe], and $\Sigma$ is the intrinsic scatter in [Si/Fe] at fixed [Fe/H]. Eq.~\ref{eq:sigma} follows a reduced chi-squared distribution with an expectation value of unity. We first calculate $\sigma^2$ assuming zero intrinsic scatter, then enforce the condition $\sigma^2$ = 1 to numerically solve for the most likely value of the intrinsic component, $\Sigma$.

We calculate the uncertainty of each measurement of $\Sigma$ using an unbiased estimator for the standard deviation of the sample variance,

\begin{equation}
s = \sqrt{ \frac{2}{N - 1} }.
\label{eq:s}
\end{equation}
Eq.~\ref{eq:s} is dependent on the number of [Si/Fe] measurements, $N$, available for each dwarf galaxy, where in general $N$ < $N_{\rm{[Fe/H]}}$ (Table~\ref{tab:obs}).  We then use $s$ to numerically solve for the corresponding upper and lower limits on $\Sigma$. We present our results for observations in Table~\ref{tab:int_scatter}.

If $\sigma^2$ = 1, $\Sigma$ = 0, whereas if $\sigma^2$ $<$ 1, the most likely value of $\Sigma$ cannot be properly determined. As long as $\sigma^2 > 1 + s$, we cannot reliably constrain the lower limit for $\Sigma$, as in the case for Draco, Fornax, and Sculptor. In these cases, $\Sigma$ is consistent with zero within one standard deviation. However, we still present the most likely range of $\Sigma$ for these dwarf galaxies in Table~\ref{tab:int_scatter}. Even if $\sigma^2 > 1$, not all values have both upper and lower limits, given the functional relationship between $\sigma^2$ and $\Sigma$. If $\sigma^2 < 1 - s$, as in the case of Peg dIrr, overestimated observational errors result in the lack of a well-determined $\Sigma$. 

Accordingly, the intrinsic scatter is highly dependent on the magnitude of $\delta$[Si/Fe]. If $\delta$[Si/Fe] is increased (decreased), $\Sigma$ decreases (increases). Thus, the measurement uncertainty in [Si/Fe] must be both accurate and precise to produce a reliable estimate of the intrinsic scatter.  The uncertainties of the $\alpha$-enhancement of the observed dwarf galaxies, determined using the \citet{Kirby2008} method of spectral synthesis of medium-resolution spectroscopy, have been validated in terms of both precision and accuracy.  \citet{Kirby2008} showed that errors in [$\alpha$/Fe] determined from medium-resolution spectroscopy remain below  0.25 dex for spectra with sufficiently high signal-to-noise ($\gtrsim$ 20 \AA). Based on duplicate observations of red giants in dwarf galaxies and the comparison to measurements of error on [$\alpha$/Fe] from high-resolution spectroscopy, \citet{Kirby2010MULTI-ELEMENT} showed that the estimated uncertainties on $\alpha$-element abundance ratios are accurate, and that the uncertainties have not been underestimated, even on an absolute scale. This indicates that our analysis sets upper limits on the true value of the intrinsic scatter for observed LG dwarf galaxies.

In most cases, we find that the most likely values of $\Sigma$ range from 0 - 0.1 dex,  except Fornax ($\Sigma =  0.14 \ \pm \ 0.02$), with upper limits of $\sim$ 0.1 - 0.17 dex. All values are consistent with zero intrinsic scatter.  We do not take into account the sampling bias associated with the observations, which primarily  target  the central, denser, and more metal-rich regions of the dwarf galaxies \citep{Kirby2011b}. 
However, the calculations of both the likelihood statistic, $\hat{L}$ (\S~\ref{sec:abnd_comp}, Eq.~\ref{eq:max_like_abnd}) and the intrinsic  scatter, $\Sigma$, take into account the limited observational sample size and observational uncertainties.  For the most likely intrinsic scatter (Table~\ref{tab:int_scatter}), larger, non-zero values result from sample size (Eq. \ref{eq:s}), excepting Fornax. Assuming that galaxies with larger sample sizes provide a more accurate measurement of the intrinsic scatter, we can conclude that it is near zero for a majority of LG dwarf galaxies.\footnote{We note that medium-resolution spectroscopy data sets necessarily have larger uncertainties than those from high-resolution spectroscopy. Although high-resolution data sets do not contain as many stars, the smaller measurement uncertainties would provide a preferable data set for quantifying the intrinsic scatter. For example, high-resolution data sets for Sculptor (\citealt{deBoer2012a}, Hill et al., in prep) and Fornax \citep{Letarte2010,deBoer2012b} could be used for this purpose.}

If the ISM is indeed homogeneous at a given time, then we should find that the intrinsic scatter at a given metallicity is nearly zero (\S~\ref{sec:mdf_width_form_time}). We conclude that, based on the near zero intrinsic scatter in [Si/Fe] at a given [Fe/H] in both simulated and observational data, including sub-grid turbulent metal diffusion in simulations produces results consistent with observations. Based on this agreement, we infer that the ISM in LG dwarf galaxies is well-mixed, to within the given scatter, throughout a galaxy's history. 

\begin{table}
\caption{Intrinsic Scatter in [$\alpha$/Fe] as a Function of [Fe/H] \label{tab:int_scatter}}
\begin{threeparttable}
\begin{tabular*}{\columnwidth}{l @{\extracolsep{\fill}} ccc}
	\toprule
    Galaxy & $\sigma^2$\tnote{a (dex)} & $\Sigma$\tnote{b} (dex) & $s$\tnote{c} (dex) \\ \midrule
    \multicolumn{4}{l}{\textbf{Simulations}} \\ \\
 	\textbf{m10q} & \nodata & 0.144 & \nodata \\
	\textbf{m10q.md} & \nodata & 0.058 & \nodata \\
	\textbf{m10v} & \nodata & 0.127 & \nodata \\
	\textbf{m10v.md} & \nodata & 0.033 & \nodata \\ \\
    \multicolumn{4}{l}{\textbf{MW dSphs}} \\ \\
    Canes Venatici I & 0.901 & $\leq$ 0.073 & 0.152\\
    Draco & 1.182 & 0.097$^{+0.036}$ & 0.098 \\
    Ursa Minor & 0.989 & $\leq$ 0.061 & 0.069 \\
    Sextans & 1.003 & 0.013$^{+0.103}$ & 0.181 \\
    Leo II & 1.061 &  0.060$^{+0.046}$ & 0.104 \\
    Sculptor & 1.173 & 0.078$^{+0.021}_{-0.023}$ & 0.079 \\
    Leo I & 0.991 & $\leq$ 0.055 & 0.056 \\ 
    Fornax & 1.417 & 0.140$^{+0.016}_{-0.015}$ & 0.059 \\ \\
    \multicolumn{4}{l}{\textbf{dIrrs}} \\ \\
    Leo A & 0.962 & $\leq$ 0.159 & 0.171 \\
    Peg dIrr & 0.667 & \nodata & 0.174 \\
    NGC 6822 & 1.045 & 0.052$^{+0.046}$ & 0.092 \\
    IC 1613 & 1.083 & 0.089$^{+0.082}$ & 0.160 \\\bottomrule
\end{tabular*}
\begin{tablenotes}
\item Note. \textemdash \ All quantities calculated in terms of [Si/Fe] vs. [Fe/H] at $z$ = 0. The ranges of $\Sigma$ represent the most likely values for observational data, although all calculations are consistent with zero.
\item[a] The initial calculation of the standard deviation of the differences between the data and the curve (Eq.~\ref{eq:sigma}), assuming zero intrinsic scatter.
\item[b] The intrinsic scatter, numerically solved for using Eq.~\ref{eq:sigma} such that $\sigma^2$ = 1. Not all values of $\Sigma$ have both upper and lower limits, owing to limits on the range of the functional relationship between $\Sigma$ and $\sigma^2$. In some cases, where $\sigma^2$ $\sim$ 1 (e.g., Ursa Minor), only an upper limit is possible to determine.
\item[c] The standard deviation of the sample variance for observational data (Eq.~\ref{eq:s}).
\end{tablenotes}
\end{threeparttable}
\end{table}

\vspace{-0.3cm}
\subsection{Comparisons to Observed Alpha-Element Abundance Patterns}
\label{sec:abnd_comp}

To determine whether the $\alpha$-element abundance ratio distributions with sub-grid metal diffusion improve statistical agreement with observations, we compare the simulations to LG dwarf galaxies. We compute a likelihood estimator, simultaneously using [Fe/H], [Mg/Fe], [Si/Fe], and [Ca/Fe], which are included in both simulated and observed data sets, as constraints. 

For each red giant star in a given observed dwarf galaxy that has a measurement for all four abundance ratios, we define a 4D Gaussian per star,

\begin{equation}
g_i = \prod_{j = 1}^{4} \frac{1}{\sigma_{[X/Y]_{ji}} \sqrt{2 \pi}} e^{-([X/Y] - [X/Y]_{ji})^2/2 \sigma_{[X/Y]_{ji}}^2},
\label{eq:fourd_gauss}
\end{equation}
where $[X/Y]$ is an abundance ratio, $\sigma_{[X/Y]}$ is the associated error, $j$ is the index for the abundance, and $i$ is the index for a star. Again, we considered only measurements with errors below 0.5 dex and [Fe/H] $>$ $-$3 dex. The 4D probability distribution for an observed dwarf galaxy with $n$ measurements is therefore,

\begin{equation}
f([X/Y]) = \frac{1}{n} \sum_{i=1}^{n} g_i([X/Y]),
\label{eq:max_like_abnd}
\end{equation}
where the prefactor is included such that the integral of the function over 4D space is unity. This can be used to estimate the log-likelihood, 

\begin{equation}
\begin{gathered}
L = \prod_{k=1}^{m} f([X/Y]_k),\\
\hat{L} = \frac{1}{m} \ln L,
\end{gathered}
\end{equation}
where $m$ is the number of star particles in a given simulated galaxy and $k$ is the index of each particle. In this case, we cite the average log-likelihood (\S~\ref{sec:mdf_comp}) to enable comparisons between different simulation/LG dwarf galaxy pairs.

We choose to include calcium abundances in the statistic determination despite a systematic offset in [Ca/Fe] in the simulations relative to observations. The simulation yields for Type II SNe result in a maximum of [Ca/Fe] = 0.1 dex at low metallicity, despite observations indicating that [Ca/Fe] $\sim$ +0.2 - +0.4 dex \citep{Venn2004,Kirby2011b}. This is likely a consequence of our assumption that there is no strong dependence of the yield on metallicity for [Fe/H] $<$ $-$2 dex in GIZMO. This is in contrast to the predictions of \citet{Nomoto2006}, which assumes metallicity dependence of the yields at low [Fe/H].

We computed the likelihoods both with (4D) and without (3D) including calcium in the product in Eq.~\ref{eq:fourd_gauss}. The results of the likelihood comparison for the abundance ratios do not differ substantially between the 3D (excluding calcium) and 4D cases in terms of the maximum likelihood matches, so we adopt the results from the 4D case, assuming that it includes additional information and subsequently provides a tighter constraint. 

The comparison between the pairs of observed and simulated galaxies results in the highest likelihoods (compared to every simulation/LG dwarf galaxy pair) of $\hat{L} = -0.456$ and $\hat{L} = -0.542$ for \textbf{m10q.md}/UMi and \textbf{m10v.md}/Leo I respectively. In comparison to the simulations without sub-grid metal diffusion, \textbf{m10v}/UMi and \textbf{m10q}/UMi have the highest likelihoods of $\hat{L} = -1.798$. 
By including TMD, we gain an increase in the statistical likelihood that the simulated and observed galaxies are similar.
Thus, we further establish (\S~\ref{sec:mdf_comp}) that Ursa Minor and Leo I are the most statistically similar dwarf galaxies in our sample to the simulations with metal diffusion, using abundance ratios in addition to MDFs. 
Including sub-grid diffusion enables galaxy simulations that provide good statistical analogues to observed dwarf galaxies. 

\vspace{-0.4cm}
\section{Environmental Effects}
\label{sec:latte}

\begin{figure}
\centering
\includegraphics[width=\columnwidth]{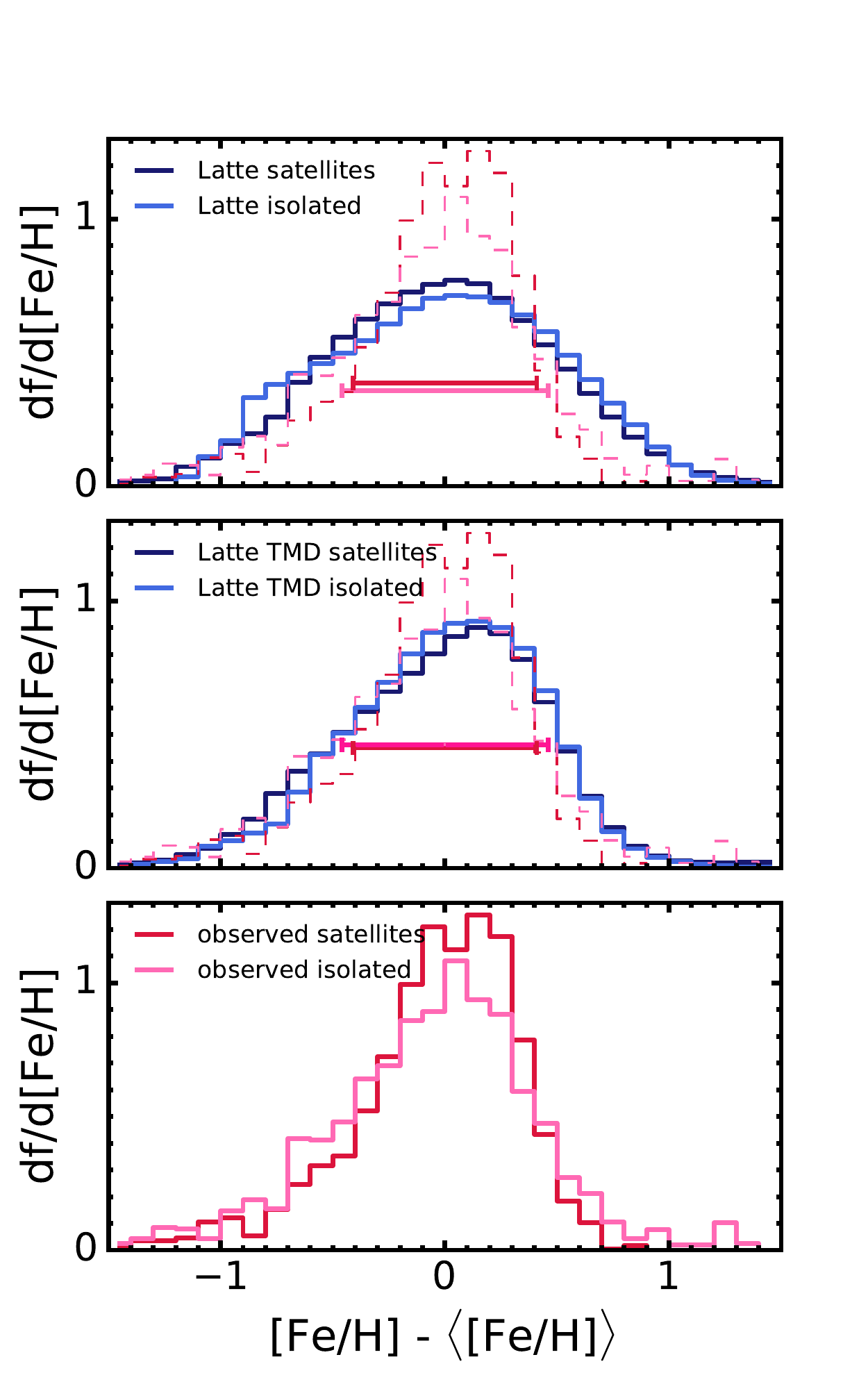}
\hspace{-0.5cm}
\caption{A comparison between average MDFs, relative to the mean metallicity for each dwarf galaxy, for both satellite and isolated dwarf galaxies with \mstar \ > 10$^{6}$ \msun at $z$ = 0. Data is shown for the Latte simulations without (\textit{top}) and with (\textit{middle}) TMD and observed LG dwarf galaxies (\textit{bottom}). The simulated MDFs are mass-weighted and smoothed to reproduce the effects of measurement uncertainty, whereas the observational MDFs are error-weighted. For comparison, we overplot average width of the observed MDFs for both satellite (red) and isolated (pink) dwarf galaxies at half maximum, as well as the full observed MDFs (dotted), on the simulated data. TMD brings the MDFs into better agreement with the observations in terms of both width and shape of the distribution, where status as a satellite or isolated galaxy does not appear to impact MDF appearance for the simulations. A  distinction between satellite and isolated galaxies may be present in the observed dwarf galaxies, although differences could be due to observational bias.\label{fig:mdf_latte_comp}}
\end{figure}

We apply the same methodology outlined in \S~\ref{sec:mdf} and \S~\ref{sec:abnd} to analyse dwarf galaxy simulations captured in the high resolution regions of the Latte simulations \citep{Wetzel2016}. Latte is run with GIZMO in MFM-mode \citep{Hopkins2015} and includes the standard FIRE-2 implementation of gas cooling, star formation, stellar feedback, and metal enrichment as summarized in \S~\ref{sec:gas_cool_sf} \citep{Hopkins2017arXiv}. We consider the \textbf{m12i} simulations, with baryonic mass resolution of 7070\ \msun \ and star particle spatial resolution of 4 pc, run with and without TMD (\S~\ref{sec:tmd}). The original \textbf{m12i} did not include TMD. We present a version of the simulation, rerun with the same initial conditions and physics, including sub-grid metal diffusion in this paper. We consider (sub)halos uncontaminated by low-resolution dark matter particles in the stellar mass range 5.5 $\times$ 10$^{5}$ \msun \ $<$ \mstar \ $<$ 9.9 $\times$ 10$^{9}$ \msun, where the lower limit is based on our metallicity convergence tests \citep{Hopkins2017arXiv}. We define ``satellite'' and ``isolated'' dwarf galaxies around the MW-like host (\mstar \ $\sim$ 6.5 $\times$ 10$^{10}$ \msun) by distance, $d_{\textrm{host}}$, with $d_{\textrm{host}}$ $<$ 300 kpc and 300 kpc $<$ $d_{\textrm{host}}$ $<$ 1 Mpc respectively, considering only simulated dwarf galaxies within the distance range of the observed LG dwarf galaxies (Table~\ref{tab:obs}). Using these criteria, we identify 10 satellite and 3 isolated dwarf galaxies for \textbf{m12i} with \mstar \ $\sim$ 7 $\times$ 10$^{5}$ - 2 $\times$ 10$^{8}$ \msun, covering a majority of the mass range spanned by the observed dwarf galaxies.

The Latte \textbf{m12i} simulation with TMD (\textbf{m12i.md}) is statistically consistent with the satellite mass function and the stellar 1D velocity dispersion of \textbf{m12i} without TMD \citep{Wetzel2016}. \textbf{m12i.md} similarly falls between the mass functions for the Milky Way and M31 (excluding the LMC, M33, and Sagittarius).  Although \textbf{m12i.md} produces fewer satellites at \mstar \ $\sim$ 10$^{6}$ \msun, this likely arises because of stochasticity in when satellites are disrupted by the host \citep{GarrisonKimmel2017}: the mass functions of isolated dwarf galaxies (d$_{\textrm{host}}$ $>$ 300 kpc), which are not affected by disruption, are nearly identical. 
Qualitatively, the stellar mass-metallicity relation between runs is in broad agreement, where the simulated dwarf galaxies agree with observations for \mstar \ $\gtrsim$ 10$^{6}$ \msun \ for a definition of the mean metallicity based on average mass fractions  (\S~\ref{sec:metal_def}). This is consistent with our previous results, where including TMD narrows the scatter in the metallicity distribution, but does not significantly change the galaxy-averaged mean metallicity. 

\vspace{-0.4cm}
\subsection{Satellite and Isolated MDFs}
\label{sec:latte_sat_iso}

\begin{figure}
\centering
\includegraphics[width=\columnwidth]{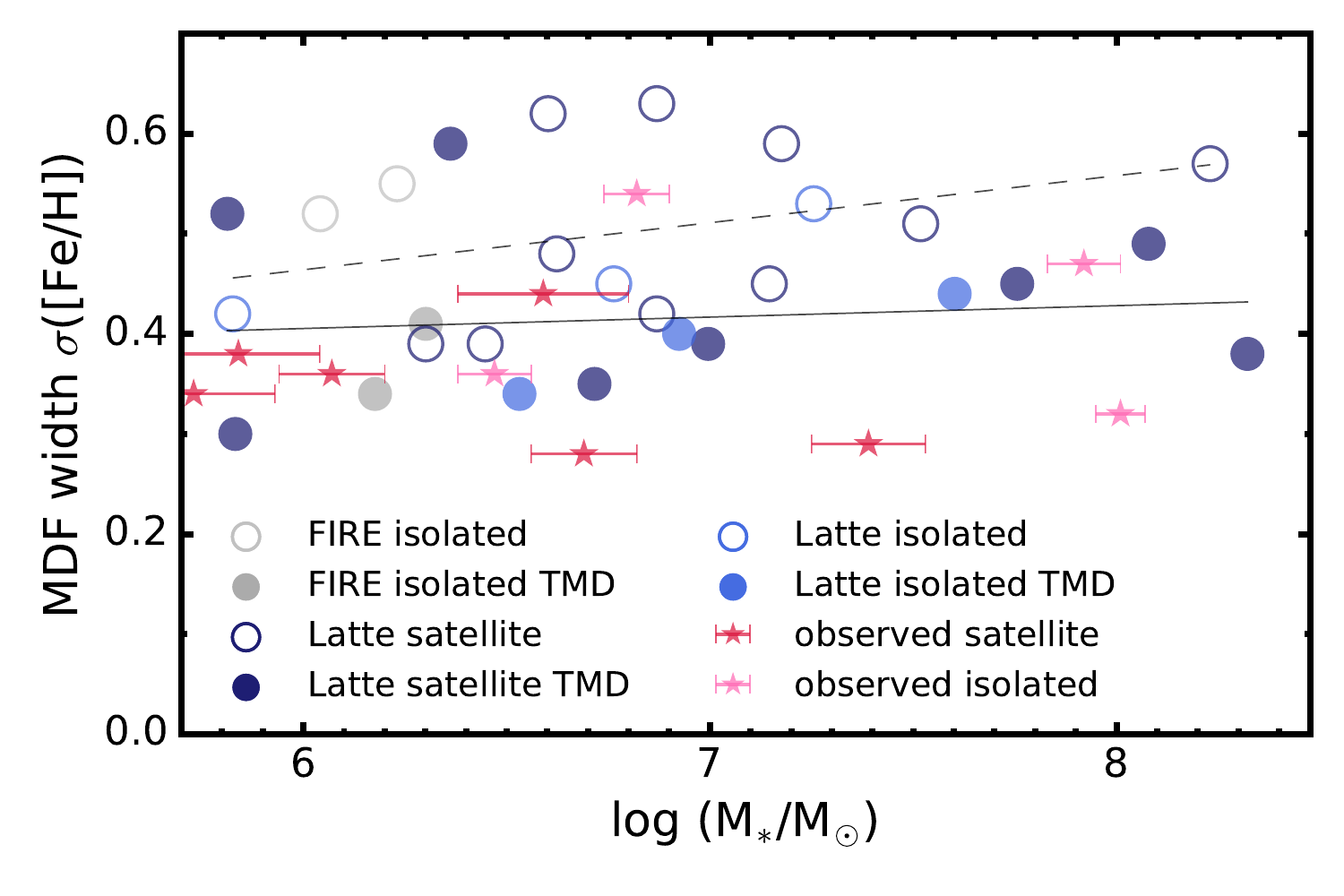}
\hspace{-0.5cm}
\caption{The MDF width as a function of stellar mass at $z$ = 0.  Best-fit lines for the trends in the simulations with stellar mass are shown for cases with (solid) and without (dashed) TMD. Latte satellite (dark blue circles) and isolated dwarf galaxies (light blue circles) with sub-grid metal diffusion, on average, have narrower MDFs across the entire stellar mass range, as compared to the run without sub-grid metal diffusion (open circles). 
For comparison, the error-corrected MDF width (Table~\ref{tab:obs}) for observed satellite (red stars) and isolated (pink stars) dwarf galaxies are shown. No apparent trend of the MDF width with stellar mass is present for all simulated dwarf galaxies with sub-grid metal diffusion. Simulations including sub-grid diffusion agree reasonably with observations, and values for the MDF width between the field FIRE and Latte dwarf galaxies are consistent with each other.
\label{fig:mdf_width_mass}}
\end{figure}

Including TMD, we observe the same narrowing effect in the widths of the MDFs for the Latte satellite and isolated dwarf galaxies as in the isolated FIRE dwarf galaxies. Figure~\ref{fig:mdf_latte_comp} shows the average MDF for dwarf galaxies with \mstar \ $>$ 10$^{6}$ \msun, excluding and including TMD. We show the MDFs for both satellite and isolated dwarf galaxy populations, where we normalize each individual dwarf galaxy MDF to its mean metallicity before averaging. The reduction in the width of the MDF is apparent, narrowing from 0.50 dex to 0.42 dex (a factor of $\sim$ 1.2) on average for combined satellite and isolated galaxy populations.  This agrees with our higher-resolution isolated dwarf galaxy simulations in \S~\ref{sec:mdf_width}. 

Figure~\ref{fig:mdf_width_mass} shows the MDF width, as a function of stellar mass, for all simulated and observed data sets. As also illustrated in Figure ~\ref{fig:mdf_latte_comp}, the Latte dwarf galaxies with TMD have narrower MDF widths on average. The values for the Latte dwarfs are consistent with those from the FIRE isolated dwarfs, albeit potentially exhibiting more scatter. Figure~\ref{fig:mdf_width_mass} emphasizes that the simulations with TMD show reasonable agreement with observational, error-corrected MDF widths (Table~\ref{tab:obs}), as opposed to runs without TMD. This is especially true for satellite dwarf galaxies. No apparent trend between MDF width and stellar mass exists for both simulated and observed dwarf galaxies. This implies that factors other than stellar mass, such as the star formation history, dictate the MDF width. In addition, it suggests that the MDF width converges in the simulations (Appendix~\ref{sec:mass_res}), since there is no mass-dependent behavior that may result from secondary resolution effects (\S~\ref{sec:int_scat_latte}).

For the simulations, both Figures~\ref{fig:mdf_latte_comp} and ~\ref{fig:mdf_width_mass} do not show any systematic differences between MDF width for satellite versus isolated galaxies. The differences between the average individual MDF widths (mass-weighted and un-smoothed) across the entire stellar mass range of simulated satellite (0.51 to 0.43 dex, or a factor of $\sim$ 1.2) and isolated (0.47 dex to 0.39 dex, or a factor of $\sim$ 1.2) galaxies without and with TMD are comparable. For both \textbf{m12i} with and without TMD, the average MDF width of isolated dwarf galaxies is narrower than the corresponding average for satellite dwarf galaxies for \mstar \ $>$ 10$^{6}$ \msun \ and $d_{\rm{host}}$ $<$ 1 Mpc. Within these constraints, there are only a few isolated dwarf galaxies in each run of \textbf{m12i}, as compared to the more numerous satellite dwarf galaxies. Expanding the sample size by incorporating a couple of uncontaminated isolated dwarf galaxies (d$_{\rm{host}}$ > 1 Mpc and  10$^{7}$ \msun \ $<$ \mstar \ $<$ 10$^{9}$ \msun) from the \textbf{m12i} simulated volume brings the average individual MDF width of isolated dwarf galaxies up to 0.50 dex and 0.42 dex, without and with TMD respectively. Thus, the average individual MDF widths between the simulated satellite and isolated dwarf galaxies are similar.

By including sub-grid metal diffusion, the average simulated MDF widths better approximate the averages of the error-corrected MDF width of individual observed galaxies (Table~\ref{tab:obs} across the entire stellar mass range for each galaxy population. The observed average satellite MDF width is 0.35 dex, whereas the observed average isolated MDF width is 0.42 dex, in comparison to 0.43 dex and 0.42 dex respectively for \textbf{m12i.md}. The observed average satellite MDF width is $\sim$ 80\% that of the observed isolated galaxies, at odds with predictions from our simulations.


Figure~\ref{fig:mdf_latte_comp} shows the analogous observational average MDFs, separated according to satellite and isolated dwarf galaxy populations.  In contrast to the simulations,  the average satellite MDF is more sharply peaked and narrow (0.41 dex) than the broader average isolated  MDF (0.46 dex), including scatter due to observational uncertainty. \citet{Kirby2013} found a less than 0.02 \% likelihood that the distributions originate from the same parent distribution, attributing the disparity to differences between star formation histories (truncated vs. extended) for satellite and isolated dwarf galaxies \citep{Mateo1998DWARFGROUP,Orban2008,Weisz2014}. Despite the similarity of the simulated satellite and isolated MDFs, both the Latte simulations and observations show no systematic difference in $\langle$[Fe/H]$\rangle$ at a given stellar mass for satellite and isolated dwarf galaxies. Further investigation of this discrepancy, such as quantifying impact of observational bias, is beyond the scope of this paper.

However,  the similarity between satellite and isolated galaxy MDFs for the same stellar mass range in the simulations indicates that TMD ultimately produces better agreement with observations. The alternate hypothesis, in which TMD mimics the impact of environmental effects (\S~\ref{sec:mdf_comp}), such as ram-pressure stripping, on the appearance of satellite MDFs, is thus excluded for the FIRE simulations.  This is further supported by a likelihood estimation comparison between the observed MDFs and the Latte simulations, in which the most similar simulated galaxies to observations were dictated solely by MDF shape, i.e., stellar mass range and star formation history, as opposed to any innate separation between isolated and dwarf galaxy MDFs. In general, the simulation suite including TMD provides better matches to the observed MDF shapes compared the case without TMD.

Figure~\ref{fig:mdf_latte_comp} illustrates that TMD ultimately better reproduces the narrowness of the observed MDFs, the truncation of stars at high metallicity, and the skew toward high metallicity. We do not observe the latter effect in the average (or any individual) MDFs for the Latte simulations without sub-grid diffusion, for both isolated and satellite galaxies. The characteristic cutoff at high metallicity in MDFs can be attributed to ram-pressure stripping and the subsequent quenching of star formation \citep{Bosler2007}. Despite this, we see similar cutoffs at high metallicity for both satellite and isolated galaxies, where isolated galaxies remain star-forming to $z$ = 0. It is unclear to what extent the cutoff is driven by environmental effects versus internal chemical evolution. Nonetheless, the cutoff appears in the simulations for \mstar \ $\gtrsim$ 10$^{7}$ \msun \ \textit{only} in the case of sub-grid metal mixing. This suggests that turbulent metal diffusion is the primary source of agreement between simulations and observations, in contrast to any significant role played by interaction with the host galaxy in the simulations.

\vspace{-0.4cm}
\subsection{Intrinsic Scatter in the Latte Simulations}
\label{sec:int_scat_latte}

\begin{figure}
\centering
 \includegraphics[width=\columnwidth]{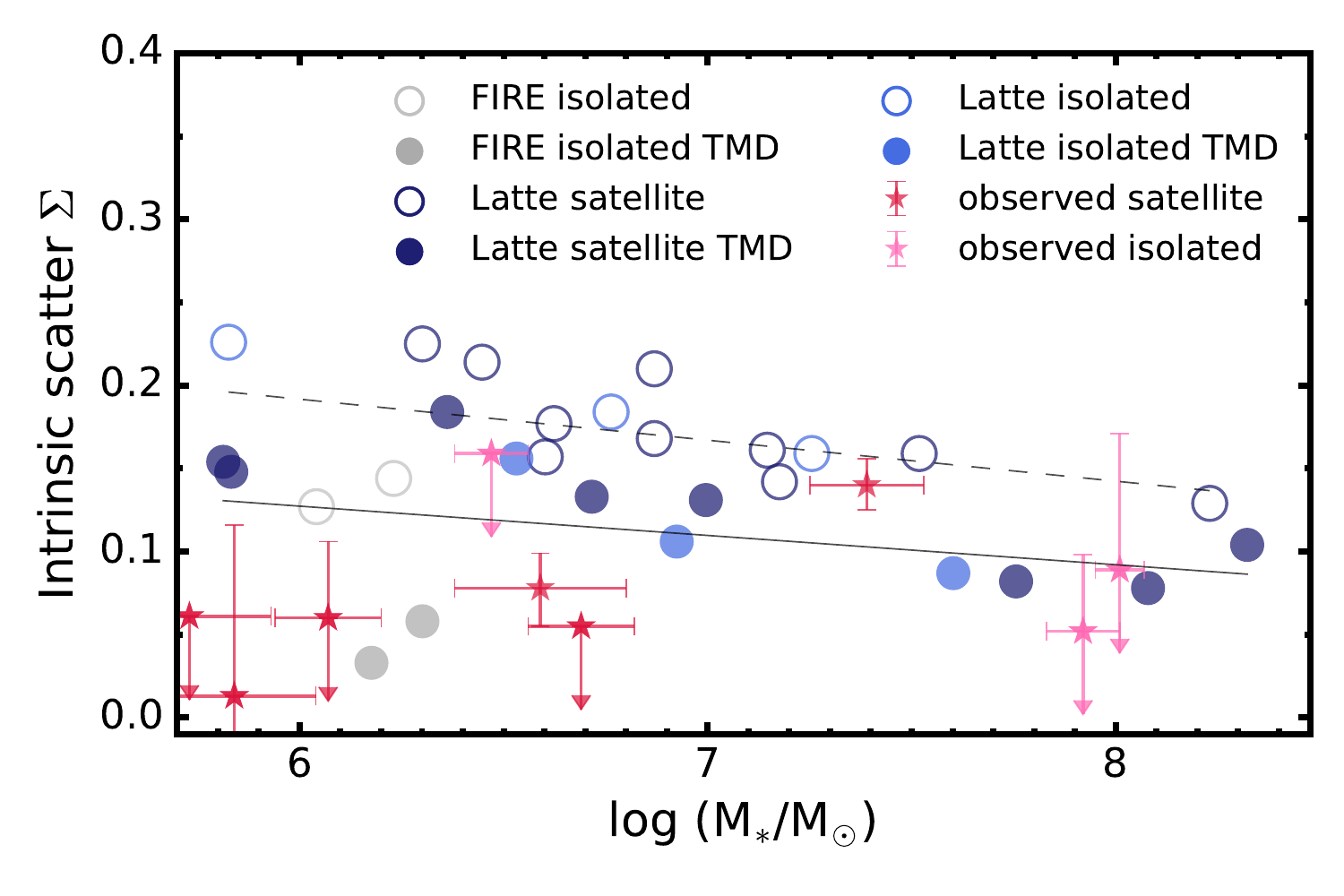}
\hspace{-0.5cm}
\caption{The intrinsic scatter in [Si/Fe] at fixed [Fe/H] as a function of stellar mass at $z$ = 0. Best-fit lines for the trends in the simulations with stellar mass are shown for cases with (solid) and without (dashed) TMD. Latte satellite (dark blue circles) and isolated dwarf galaxies (light blue circles)  with metal diffusion have lower intrinsic scatter across the stellar mass range \mstar \ $\sim$ 10$^{6}$ - 10$^{8}$ 
\msun \ as compared to the runs without sub-grid metal diffusion (open circles).  The reduction in the intrinsic scatter upon including TMD is less pronounced for the Latte simulations as compared to the FIRE isolated dwarfs (grey circles) at a given stellar mass for \mstar \ $\lesssim$ 10$^{6.5}$ \msun, owing to resolution effects.  For comparison, upper limits of the the most likely values of the intrinsic scatter for the for observed satellite (red stars) and isolated (pink stars) dwarf galaxies are shown. The higher-resolution isolated dwarf galaxy FIRE simulations, for which the effects of TMD are more important, are in better agreement with MW dSphs, which tend to have better measurements of $\alpha$-enhancement than LG dIrrs. 
\label{fig:int_scat_latte}}
\end{figure}

We calculate the intrinsic scatter for Latte dwarf galaxies, with and without TMD.  The average reduction in the intrinsic scatter for galaxies with \mstar \ $>$ 10$^{6}$ \msun \ is from 0.16 to 0.12 dex,  or a factor of $\sim$ 1.1. This is comparable to the reduction in the width of the MDF and the average reduction factor for the FIRE isolated dwarfs ($\sim$ 1.2).  However, the reduction in intrinsic scatter including TMD at a given stellar mass is less pronounced (Figure~\ref{fig:int_scat_latte}) in the lower-resolution Latte simulations, as compared to the higher-resolution isolated FIRE dwarf galaxies.  In addition, the Latte dwarf galaxies have larger intrinsic scatter overall as compared to the FIRE isolated dwarf galaxies at \mstar \ $\lesssim$ 10$^{6.5}$ \msun.

This is likely caused by secondary resolution effects (Appendix~\ref{sec:mass_res}). Numerically enhanced burstiness \citep{Hopkins2017arXiv} results in stronger periods of inhomogeneity for the less well-resolved low-mass  (\mstar \ $\lesssim$ 10$^{6.5}$ \msun) Latte dwarf galaxies. By definition, the intrinsic scatter is more sensitive to inhomogeneities than the MDF width, which is more sensitive to the long-term star formation history, artificially increasing the dispersion in the $\alpha$-element abundances at fixed metallicity. 
For this reason, low-mass  Latte dwarf galaxies do not agree with most observations of the MW dSphs with near-zero intrinsic scatter ($\lesssim$ 0.1 dex), which have more reliable measurements of $\alpha$-enhancement than LG dIrrs (Figure ~\ref{fig:int_scat_latte}). The impact of turbulent metal diffusion is more important for higher-resolution simulations, in which the effects of numerical noise become more pronounced (Appendix~\ref{sec:imf},~\ref{sec:metal_inject}), in terms of bringing simulations of dwarf galaxies into agreement with observations.  


\vspace{-0.3cm}
\section{Summary \& Discussion}

We have examined the metallicity distribution functions and enrichment histories, including [Fe/H] $\alpha$-element abundances, of dwarf galaxies using the FIRE-2 cosmological simulations of dwarf galaxies, to investigate the chemical enrichment histories of dwarf galaxies. We have shown that turbulent metal diffusion, at levels suggested by converged simulations, is necessary to include in Lagrangian, hydrodynamical simulations of the formation and evolution of dwarf galaxies to obtain realistic predictions of chemical evolution. 

Contrary to a majority past studies of the chemical properties of simulated dwarf galaxies, we have successfully modelled full abundance distributions, in addition to global properties such as the stellar mass-metallicity relation. As a caveat, we note the presence of an offset in the normalization of our stellar mass-metallicity relation compared with observations of low-mass dwarf galaxies. This is likely caused by systematic effects, such as our choice of the Type Ia SNe delay time distribution. However, this does not impact properties such as the MDF width and the intrinsic scatter in [$\alpha$/Fe] at fixed [Fe/H].

Through statistical comparison of FIRE simulations to observations of LG dwarf galaxies, we have demonstrated that simulations including TMD are in agreement with the width of the MDF and the intrinsic scatter in [$\alpha$/Fe] vs. [Fe/H]. For both the MDF and the  $\alpha$-element abundance ratios, a reduction in the scatter occurs compared to simulations without TMD, as well as a reduction in numerical artifacts such as  star particles with [Fe/H] $<$ $-$3 and envelopes corresponding to Type Ia/Type II SNe yields. 

The same effects are present in both satellite and isolated dwarf galaxies from the Latte simulation, albeit the reduction in [$\alpha$/Fe] vs. [Fe/H] scatter is subdued owing to resolution effects. Most significantly, we find that a distinction between satellite and isolated dwarf galaxies does not factor into our conclusions on the agreement between simulations and observations in terms of the shape and width of the MDF and dispersion in the $\alpha$-element abundance ratio distributions. Just as simulated and isolated dwarf galaxies are similar in these quantities, all dwarf galaxies, both observed and simulated, lie on the same stellar mass-metallicity relation regardless of environment or star formation history \citep{Skillman1989,Kirby2013}. In addition, both satellite and isolated dwarf galaxies primarily form as dispersion-dominated systems regardless of current proximity to the host \citep{Wheeler2017,Kirby2017}. This also poses a challenge to the traditional separation between dwarf galaxy populations. Our work suggests that galactic chemical evolution depends predominantly on the stellar mass of dwarf galaxies.

We analyse realistic  galaxy evolution and formation simulations,  taking into account a multi-phase ISM, explicit stellar feedback, and the impact of cosmological accretion. Our analysis serves as a robust confirmation of previous work  done in the case of idealized, non-cosmological simulations. We have illustrated that a sub-grid turbulent diffusion model, owing to the physically-motivated nature of the implementation and its ability to match observations, is a valid alternative to methods such as a smoothed-metallicity scheme \citep{Wiersma2009,Revaz2016}.  
Similar to \citet{Williamson2016},  we find that the strength of mixing due to turbulent diffusion is stable against variations in the diffusion coefficient  within an order of magnitude above a minimum diffusion strength (Appendix~\ref{sec:diff_coeff_stability}).

In addition, for the first time, we have presented  an explicit calculation of the  intrinsic scatter from medium-resolution spectroscopy of 12 LG dwarf galaxies. Previous studies of metal poor Galactic stars \citep{Carretta2002,Cayrel2004, Arnone2005} similarly found near-zero intrinsic scatter for $\alpha$-elements at fixed metallicity. Based on $\alpha$-element abundance ratios (see \citealt{HiraiSaitoh2017} for an analysis based on barium abundances), we conclude that the timescale for metal mixing is shorter than the typical dynamical timescale for dwarf galaxies. This is evidenced by the homogeneity of the ISM, as implied by the near zero intrinsic scatter for LG dwarf galaxies. 

The implication of a well-mixed ISM for one-zone chemical evolution models (e.g., \citealt{Lanfranchi2003,Lanfranchi2004TheGalaxies,Lanfranchi2007,Lanfranchi2010,Lanfranchi2006}) is that 3D hydrodynamical models \citep{Mori2002,Revaz2009,Sawala2010} may not be necessary to relax the instantaneous mixing approximation, since the cool gas of the ISM becomes homogeneous within approximately a dynamical time for dwarf galaxies. This is in contrast to previous studies \citep{Marcolini2008} that investigated the effects of inhomogeneous pollution by SNe on chemical properties in 3D hydrodynamical simulations of isolated dSphs. However, \citet{Marcolini2008} did not include a prescription for sub-grid metal mixing, which washes out temporary inhomogeneities in the ISM. Based on our analysis, one-zone approximations in chemical evolution models may be appropriate for dwarf galaxies.

Turbulent metal diffusion is important for accuracy in modelling the ISM as well as processes relevant for chemical evolution. The inclusion or exclusion of TMD  will therefore influence predictions drawn from simulated chemical abundances. For example, \citet{Bonaca2017} justified the use of the Latte  \textbf{m12i} primary halo as a Milky Way analogue for comparison to data from  \textit{Gaia} Data Release 1 \citep{Gaia2016a,Gaia2016b} using the width of the MDF including TMD.  The authors then drew inferences on the hierarchical formation of the Galaxy and its halo structure based on the simulation. 

Including turbulent metal diffusion can enable the use of simulations in detailed, chemical abundance based investigations of galaxy formation and evolution. 



\vspace{-0.3cm}
\section*{Acknowledgements}


IE would like to thank the anonymous referee, in addition to Shea Garrison-Kimmel, Matthew Orr, and Denise Schmitz, for helpful comments that improved this paper. Numerical calculations were run on the Caltech computing cluster ``Zwicky'' (NSF MRI award \#PHY-0960291) and allocation TG-AST130039 granted by the Extreme Science and Engineering Discovery Environment (XSEDE) supported by the NSF. IE was supported by Caltech funds, in part through the Caltech Earle C. Anthony Fellowship, and a Ford Foundation Predoctoral Fellowship. AW was supported by a Caltech-Carnegie Fellowship, in part through the Moore Center for Theoretical Cosmology and Physics at Caltech, and by NASA through grants HST-GO-14734 and HST-AR-15057 from STScI. ENK was supported by NSF Grant AST-1614081. Support for PFH was provided by an Alfred P. Sloan Research Fellowship, NASA ATP Grant NNX14AH35G, NSF Collaborative Research Grant \#1411920, and CAREER grant \#1455342. CW was supported by the Lee A. DuBridge Postdoctoral Scholarship in Astrophysics. DK was supported by NSF grant AST-1715101 and the Cottrell Scholar Award from the Research Corporation for Science Advancement. CAFG was supported by NSF through grants AST-1412836 and AST-1517491, by NASA through grant NNX15AB22G, and by STScI through grant HST-AR-14293.001-A. EQ was supported in part by a Simons Investigator Award from the Simons Foundation and NSF grant AST-1715070.



\bibliographystyle{mnras}
\bibliography{Mendeley}



\appendix \label{sec:appendix}

\vspace{-0.4cm}
\section{Stochastic IMF Sampling}\label{sec:imf}

Here,  we note that additional scatter can be introduced into the MDFs and abundances as a consequence of stochastic IMF sampling at sufficiently high resolutions, such that a star particle no longer approximates a single stellar population. Estimates from \citet{Revaz2016} for SPH methods suggest that the single stellar population approximation no longer holds for star particles with \mstar \ $<$ 1000 \msun,  which includes the  standard FIRE dwarf galaxies with mass resolution of $\sim$ 250 \msun.  

We do indeed expect that the IMF is not being individually well-sampled at these masses. However, we anticipate that this is more of an issue in ultra-faint dwarf galaxies,  which have sufficiently low stellar masses such that the star formation history of the galaxy is substantially impacted by a single SN. In contrast, a majority of the FIRE dwarf galaxies have \mstar \ $\gtrsim$ 10$^{6}$ \msun, such that the effects of individual SNe on galaxy-scale properties are negligible, or weak at most (Su et al., in prep.).

We expect that the  additional numerical scatter introduced by IMF sampling is offset by the inclusion of IMF-averaged yields \citep{Iwamoto1999, Nomoto2006}. Ultimately, we are concerned with total metallicity  distributions and the behaviour of $\alpha$-element abundance ratios as a function of metallicity, as opposed to detailed abundance patterns.  In the latter case,  the effects of IMF sampling would be more pronounced, and would  require proper quantification. Particularly in the case of metal diffusion, we  do not anticipate that stochastic IMF sampling  will have a significant effect on the MDFs and abundance ratios, since metal diffusion drives the metallicity of a star particle toward the average metallicity of the galaxy, and so can reduce artificial noise from a number of different sources. 

Sub-grid metal mixing, which we have argued is  necessary to include in  hydrodynamical simulations of galactic chemical evolution, has interesting implications  for modelling individual abundance patterns. That is, the presence of sub-grid diffusion or the lack thereof dictates the amount of information that can be extracted from detailed abundance patterns. With the inclusion of metal diffusion, only the first few SN explosions individually affect a galaxy, after which metals are quickly homogenized. 

Overall, this study will be informative for addressing how much scatter can be attributed to IMF sampling, and to what degree numerical artifacts can be reduced via sub-grid metal mixing. 

\vspace{-0.4cm}
\section{Metal deposition}\label{sec:metal_inject}

\begin{figure}
\centering
\includegraphics[width=\columnwidth]{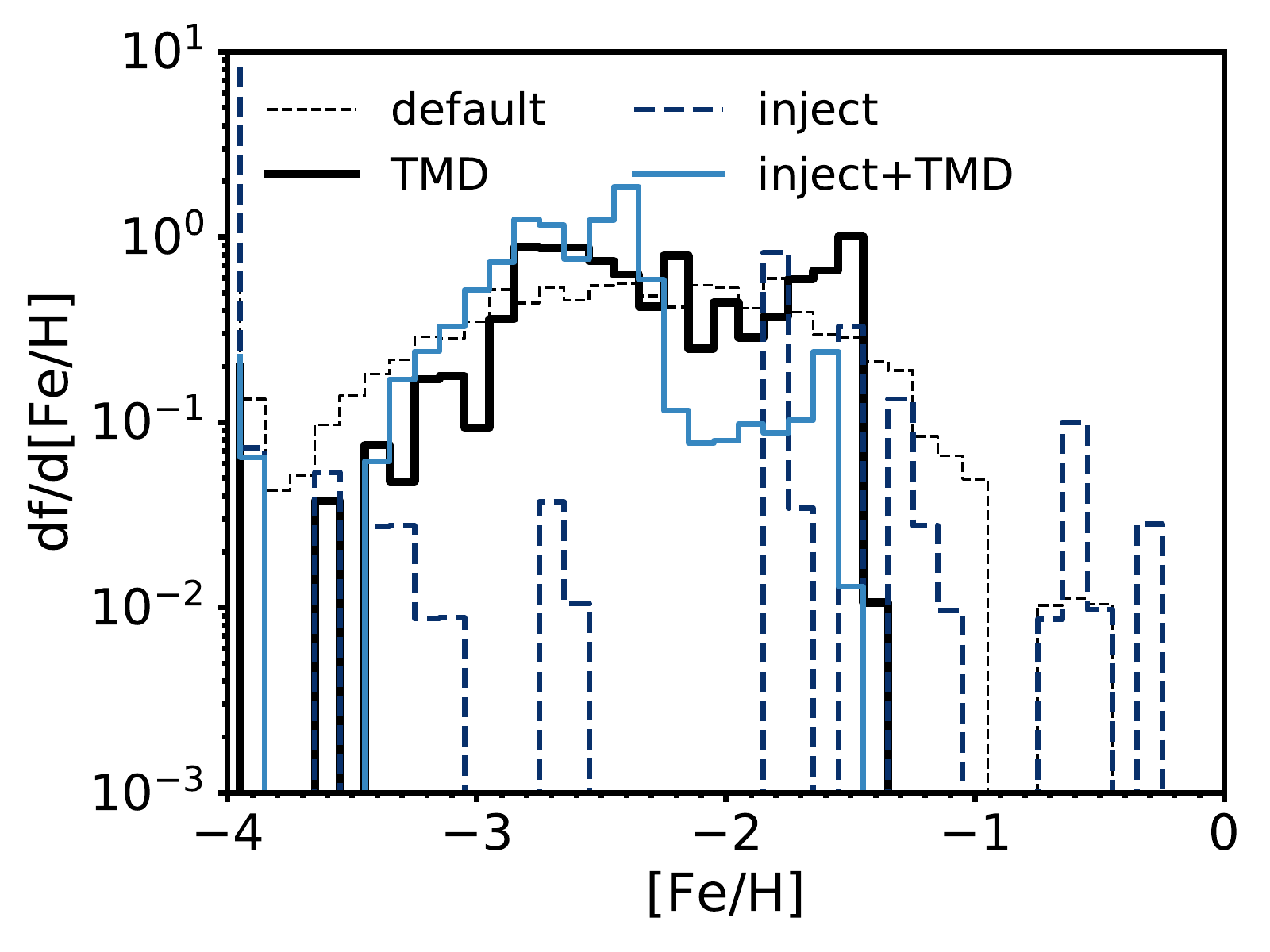}
\hspace{-0.5cm}
\caption{Stellar-mass weighted metallicity distribution functions at $z$ = 0 with differing combinations of metal injection methods and including (solid lines) and excluding (dashed lines) turbulent metal diffusion. The simulations are run at a comparatively lower mass resolution of 2100 \msun, with the same initial conditions as \textbf{m10q}. The identifier ``inject'' indicates that all metals are injected into a single particle per enrichment event (blue lines), as opposed to the standard metal injection scheme (black lines), whereas ``TMD'' indicates the presence of turbulent metal diffusion. The metallicity distribution functions are plotted on a log scale to emphasize the behaviour at the tail of the distributions. The inclusion of diffusion offsets the effects of the unphysical metal injection scheme, implying that diffusion stabilizes results against differences in the specific numerical implementation of metal deposition. \label{fig:inject_var}}
\end{figure}

Altering the metal deposition algorithm  could potentially impact the  MDF and abundances by introducing an additional source of numerical scatter.  For example, gas particles with different distances from the ``exploding'' star particle may receive varying amounts of metals, a fixed number of particles may be injected with metals, or a fixed volume surrounding such an exploding star particle may be enriched.  A  more realistic metal deposition prescription , or one in which the metals are dispersed more uniformly, will  intuitively introduce less scatter into  chemical evolution observables as compared to, e.g., the injection of all metals from a SN into a single gas particle.

To  illustrate the robustness of our results with respect to the details of mechanical feedback when including diffusion, we compare the unphysical, extreme case of single-particle metal injection to that of the standard FIRE implementation. In the standard method, an effective neighbour number is determined based on a kernel function and search radius within a sphere defined by these quantities, such that the resulting distribution of ejecta is isotropic, including regions with highly disordered gas particle positions \citep{Hopkins2017arXiv}. 

Figure~\ref{fig:inject_var} shows the logarithmic MDFs of four simulations run at a comparatively lower mass resolution of 2100 \msun \ for \textbf{m10q}. As expected, the MDF for the case of single-particle injection, without TMD, exhibits an unrealistically large scatter characterized by a series of disconnected peaks. In this prescription, some star particles are never enriched, whereas others are enriched to high values of metallicity ([Fe/H] $>$  $-$0.5) that are not observed in LG dwarf galaxies. 

The inclusion of TMD brings the single-particle injection method into better approximate agreement with standard FIRE MDFs with and without sub-grid diffusion in terms of the shape of the distribution (Figure~\ref{fig:inject_var}). All simulations presented in Figure~\ref{fig:inject_var} have approximately the same stellar mass at z = 0 (\mstar \ $\sim$ 1.2 - 1.7 $\times$ 10$^{6}$ \msun). The same narrowing effect of the MDF is exhibited, with a reduction in the width from $\sigma$[Fe/H] = 0.41 dex to 0.30 dex. This accompanies a significant shift of the average metallicity from -1.51 dex to -2.49 dex (as compared to average metallicity values of [Fe/H] = -2.18 and -2.15 for the standard metal deposition scheme with and without  sub-grid diffusion). The narrowing effect is not as pronounced in the  runs with standard metal deposition, with $\sigma$[Fe/H] =  0.50 and 0.46 dex with and without diffusion respectively, most likely owing to lower mass resolution (2100 \msun \ as compared to 250 \msun).  

Thus, we conclude that the presence of diffusion stabilizes the MDFs and abundances against differences in the specific numerical implementation of metal deposition.

\vspace{-0.4cm}
\section{Diffusion coefficient calibration} 
\label{sec:diff_coeff}

\begin{figure}
\centering
\includegraphics[width=\columnwidth]{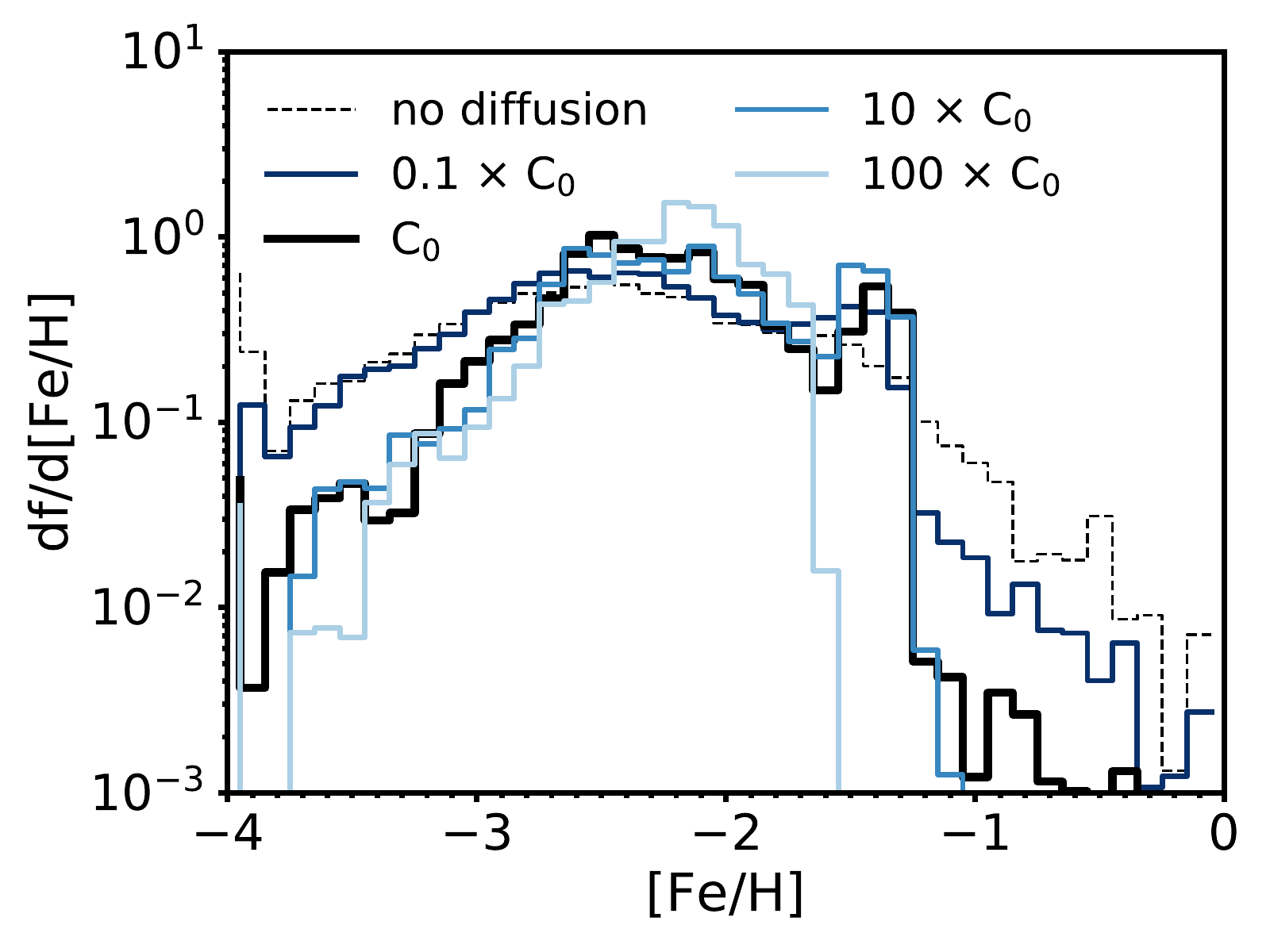}
\hspace{-0.5cm}
\caption{Stellar mass-weighted metallicity distribution functions at z = 0 with varied diffusion coefficients (blue lines), altering the overall diffusion strength, for runs of \textbf{m10q}. \textbf{m10q} without sub-grid metal diffusion is shown for reference (dashed black line). The metallicity distribution functions are plotted on a log scale to emphasize the behaviour at the tails of the distribution. 10 $\times$ C$_{0}$ represents a diffusion strength a factor of ten larger than the default value of C$_0$ $\approx$ 0.003 (thick black line; \S~\ref{sec:diff_coeff_stability}), and so on. C$_0$ represents the minimum diffusion strength for significant sub-grid metal mixing to occur. 
The lack of a significant difference between diffusion effects for different values of the coefficient beyond our fiducial range suggests that MDF predictions are not extremely sensitive to the exact adopted value of the free parameter above the minimum diffusivity. However, in the case of the a diffusion strength larger by two orders of magnitude (100 $\times$ C$_0$), the effects of over-mixing become apparent. \label{fig:coeff_var}}
\end{figure}

While the implementation of turbulent metal diffusion is more physical, given the nature of turbulence in the ISM, the diffusion coefficient must be calibrated independently for different numerical methods, owing to the different ``effective resolution scale'' of the turbulent cascade associated with each simulation. For example, the relevant scale at standard FIRE dwarf resolution ($\sim$ 250 M$_{\odot}$) will differ compared to other simulations, since a majority of the mixing is resolved. \citet{Colbrook2017} tested the FIRE implementation of turbulent metal diffusion in idealized, converged turbulent box simulations, verifying that the prescription is valid with $C$ $\approx$ 0.05, where $C_0$ = $\sqrt{2} C^2$, given our definitions of \textbf{h} and \textbf{S}. This is further confirmed by a more comprehensive study by Rennehan et al. (in prep.), which similarly indicates that $C$ $\approx$ 0.03 - 0.05.  Thus, a proper calibration  of the coefficient is necessary to ensure that sub-grid metal diffusion approximates a  physical process at a given resolution-scale. We emphasize through the numerical tests shown in \S~\ref{sec:diff_coeff_stability} that the method is not especially sensitive to the exact value of the coefficient for a given calibration.

\vspace{-0.3cm}
\section{Robustness of Results with respect to the Diffusion Coefficient}
\label{sec:diff_coeff_stability}

A potential drawback of the sub-grid turbulent metal diffusion implementation is possibly unphysical amounts of diffusion. To address the possibility of significant over-mixing, we varied the diffusion coefficient, $C_0$ (\S~\ref{sec:tmd}), to test the impact on the strength of metal diffusion on the MDF. 

As illustrated in Figure~\ref{fig:coeff_var}, an increase in the coefficient by an order of magnitude for runs of \textbf{m10q} does not significantly change the appearance of the MDF above a minimum diffusivity. A reduction of the diffusion strength by an order of magnitude (0.1 $\times$ C$_0$), relative to the standard adopted diffusion strength ($C_0$ = 0.003), results in a broader MDF (0.50 dex). Although the detailed distribution at the high-metallicity tail changes for a diffusion strength of 0.1 $\times$ C$_0$, the low-metallicity tail and main body of the MDF are comparable to the case without any sub-grid metal diffusion.
We calibrate the value of $C_0$ such that it corresponds to the minimum diffusion strength that results in significant sub-grid metal mixing. Above $C_0$ = 0.003, increasing the diffusion strength results in some reduction in the number of stars present in both the high- and low-metallicity tails. Despite the increase in the diffusion coefficient, the widths of the distributions  remain approximately constant at 0.44 dex and 0.46 dex for diffusion coefficients of C$_0$ and 10 $\times$ C$_0$ respectively. However, in the case of  an increase in the coefficient by two orders of magnitude (100 $\times$ C$_0$), the effects of over-mixing become apparent with a reduction in the width to 0.29 dex, as well as in the complete absence of the  tails of the distribution.
 
This suggests that MDF predictions are not extremely sensitive to the exact diffusion coefficients above a minimum diffusivity within an order of magnitude, as long as some degree of sub-grid turbulent mixing is present.  Together, the insensitivity of the MDF to the diffusion coefficient and the narrowing of the MDF relative to runs without TMD (\S~\ref{sec:mdf_width}) imply that the timescale for metal mixing is shorter than a dynamical time for dwarf galaxies. As we show explicitly in \S~\ref{sec:mdf_width_form_time}, with any given  burst of star formation, we can assume that the ISM is well-mixed and nearly homogeneous, resulting in a stellar population born with approximately the same metallicity.

Previous work by \citet{Williamson2016} involving idealized, non-cosmological simulations of dwarf galaxies came to a similar conclusion concerning the robustness of the diffusion strength relative to the diffusion coefficient. However, this work does not include the infall of pristine gas, galaxy interactions, galaxy evolution through cosmological time, the initialization of star particles at low-metallicity at early times (to approximate the Population III to Population II transition), and self-regulated star formation, which are essential for an accurate model of the chemodynamical evolution of a galaxy. Thus, we confirm that the addition of diffusion still contributes to the robustness of results in the case of more realistic galaxy evolution simulations.

\vspace{-0.4cm}
\section{Mass resolution}\label{sec:mass_res}

We consider the impact of mass resolution on the width of the MDF and the intrinsic scatter in [$\alpha$/Fe] vs. [Fe/H]. The star particle mass resolution for the isolated FIRE dwarfs is 250 \msun \, whereas the Latte dwarf galaxies are at lower mass resolution (7070 \msun), and thus resolution effects may factor into our comparisons between isolated FIRE dwarf galaxies and dwarf galaxies from the Latte simulations.

First, we establish that most results, such as the stellar mass and star formation rates, converge with mass resolution after the Toomre scale (i.e. the largest self-gravitating structures) is resolved \citep{Hopkins2011,Hopkins2017arXiv}. The metallicity and burstiness of dwarf galaxy star formation histories converge to within $\sim$ 10 \% (to $\sim$ 20\% maximum) differences  for (1) $N$ $\gtrsim$ 100, where $N$ is the number of star particles in a galaxy, such that the self-enrichment history is well-sampled, and (2) star particle mass $\lesssim$ 10$^{4}$ \msun \ for M$_{\textrm{halo}}$ $\sim$ 10$^{10}$ \msun, such that numerically enhanced burstiness does not occur \citep{Hopkins2017arXiv}. 

\begin{figure}
\centering
\includegraphics[width=\columnwidth]{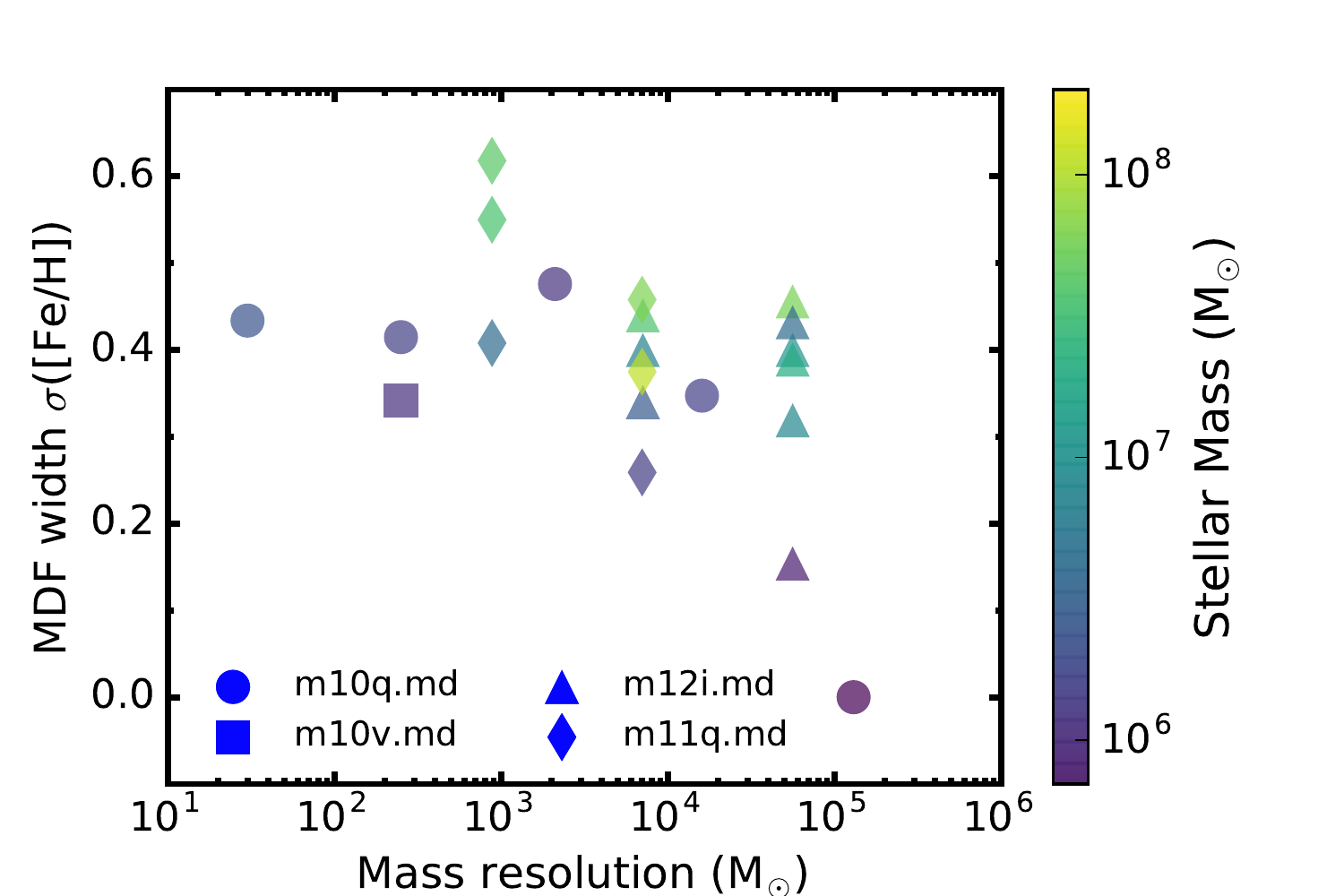}
\hspace{-0.5cm}
\caption{MDF width as a function of mass resolution at $z$ = 0 for the FIRE isolated dwarfs \textbf{m10v} and \textbf{m10q}. Isolated dwarf galaxy subhalos of both \textbf{m11q}, a primary host with LMC mass, and \textbf{m12i} are also shown. We consider runs with TMD, since runs without TMD have star particles with more improbable metallicities and have comparatively enhanced numerical noise in the MDF. The data is colour coded according to the stellar mass of the galaxy, to account for the potential of associated variation in the MDF width.  No systematic trend of the MDF width exits with mass resolution for baryonic particle masses $\lesssim$ 10$^{4}$ \msun.
\label{fig:mass_res_mdf}}
\end{figure}

Figure~\ref{fig:mass_res_mdf} illustrates the results of mass resolution tests for the MDF width. Data from the FIRE isolated dwarfs, including an ultra high-resolution (30 \msun) run of \textbf{m10q} (Wheeler et al., in prep), in addition to isolated dwarf galaxies in the zoom-in region of both \textbf{m11q}, a LMC mass halo, and \textbf{m12i} are shown. We consider only isolated dwarfs for the resolution test to account for differences in the disruption of satellites by the host (\S~\ref{sec:latte}). All runs include TMD, where runs without TMD contain star particles with more improbable metallicites and have comparatively enhanced numerical noise in the MDF. 

Although the MDF width exhibits some variation for the isolated dwarf galaxies in simulations with a primary host, it can be attributed to the typical scatter  (standard deviation $\lesssim$ 0.1 dex) at a given stellar mass expected from stochastic effects. Nonetheless, the mean MDF width appears to systematically decrease for mass resolution $\gtrsim$ 10$^{4}$ \msun. We note that the lowest resolution run, at baryonic particle mass of 160,000 \msun, has a near zero MDF width likely owing to a truncated SFH caused by numerically enhanced burstiness, a secondary resolution effect. Excluding the lowest resolution run and taking into account the typical scatter at a given stellar mass, we conclude that the MDF width of the FIRE isolated dwarfs and Latte satellite and isolated dwarf galaxies are converged for mass resolution $\lesssim$ 10$^{5}$ \msun. 

\begin{figure}
\centering
\includegraphics[width=\columnwidth]{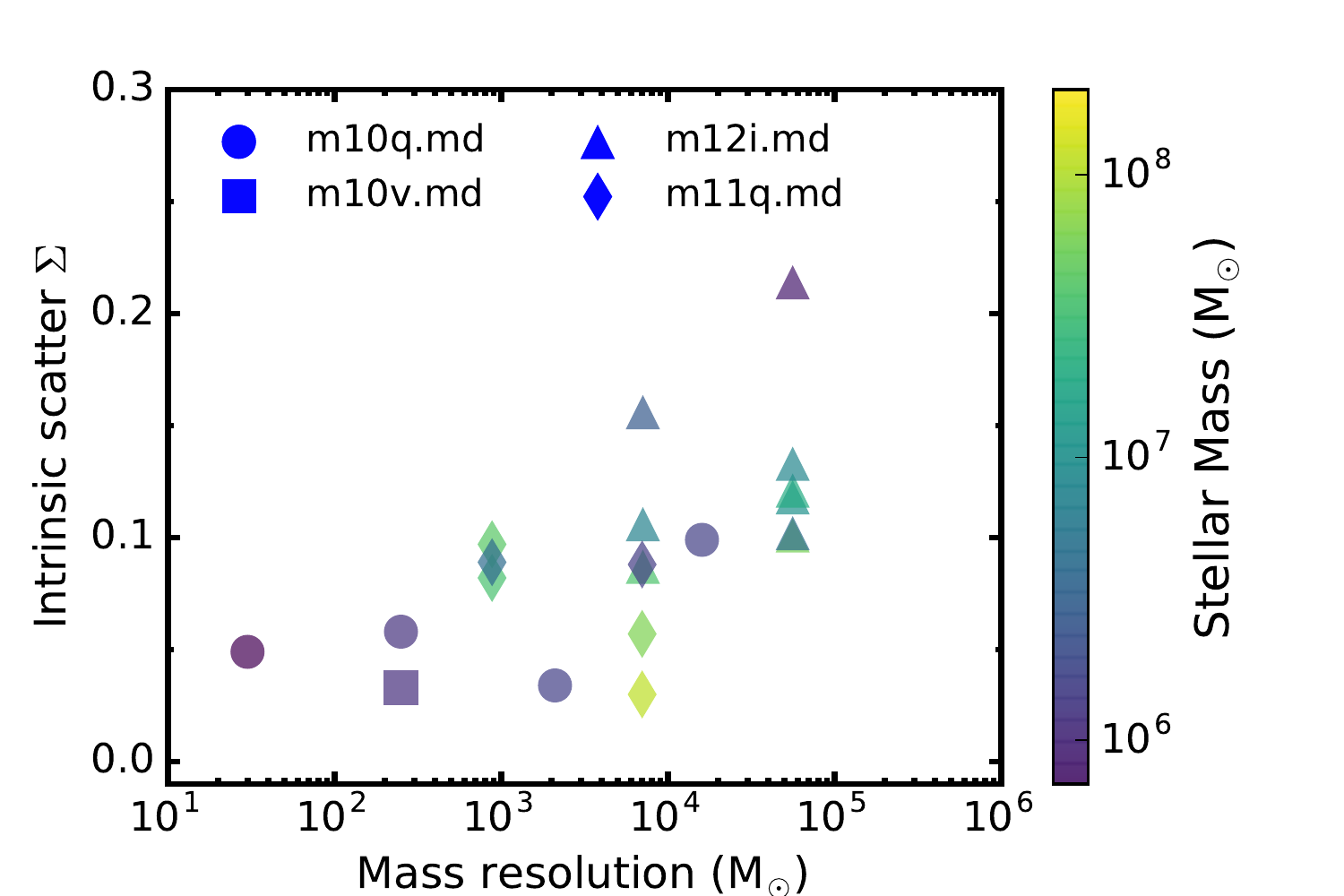}
\hspace{-0.5cm}
\caption{Intrinsic scatter in [Si/Fe] at fixed [Fe/H] as a function of mass resolution at $z$ = 0. All other aspects of the figure are identical to Figure~\ref{fig:mass_res_mdf}. In contrast to the MDF width, the intrinsic scatter, $\Sigma$, appears to converge for baryonic particle masses $\lesssim$ 10$^{3}$ \msun, taking into account the full range of stellar masses. Considering only high-mass (better resolved) dwarf galaxies, the intrinsic scatter converges for mass resolution $\lesssim$ 10$^{4}$ \msun.
Thus, we expect $\Sigma$ to be larger on average for the Latte simulations as compared to the higher-resolution FIRE isolated dwarf galaxies at low stellar masses (\mstar \ $\lesssim$ 10$^{6.5}$ \msun. \label{fig:mass_res_int_scat}}
\end{figure}

Similarly, Figure~\ref{fig:mass_res_int_scat} shows mass resolution tests for the intrinsic scatter in [Si/Fe] at fixed [Fe/H]. In contrast to the MDF width, the intrinsic scatter increases with mass resolution $\gtrsim$ 10$^{3}$ \msun \ for the lowest mass simulated dwarf galaxies (\mstar \ $\lesssim$ 10$^{6.5}$ \msun). This is due to secondary resolution effects resulting in numerically enhanced burstiness, as discussed in \S~\ref{sec:int_scat_latte}. We quantify the degree of burstiness as a the fraction of the total $z$ = 0 stellar mass formed when the star formation rate averaged over short timescales ($\sim$ 10 Myr) exceeds (by a factor of 1.5) the average star formation rate averaged over longer timescales ($\sim$ 100 Myr). We find that burstiness converges for baryonic particle mass resolution $\lesssim$ 10$^{3}$ \msun.
For mass resolution $\gtrsim$ 10$^{3}$, burstiness in dwarf galaxies increases with decreasing mass resolution, while simultaneously exhibiting mass-dependent behaviour, such that lower-mass dwarf galaxies are burstier. 

Hence, the intrinsic scatter in Latte dwarf galaxies with baryonic particle mass 7070 \msun increases for less well-resolved, lower-mass dwarf galaxies. The intrinsic scatter converges at higher mass resolution than the MDF width, because the burstiness is more sensitive to mass resolution than the long-term ($\sim$ 10 Gyr) star formation history.  Since we cannot separate resolution effects from stochastic variation in the intrinsic scatter at a given stellar mass for these dwarf galaxies, we do not quantify the typical variation expected in the intrinsic scatter, $\Sigma$, as in the case of the MDF width, $\sigma$([Fe/H]). Considering only the better resolved, high-mass dwarf galaxies (\mstar \ $\gtrsim$ 10$^{6.5}$ \msun), no trend exists in the intrinsic scatter with baryonic particle mass resolution for $\lesssim$ 10$^{4}$ \msun \ for both the FIRE isolated and Latte dwarf galaxies.


\bsp	
\label{lastpage}
\end{document}